\documentclass[sigconf]{acmart}
\AtBeginDocument{%
  }

\copyrightyear{2025}
\acmYear{2025}
\setcopyright{acmlicensed}\acmConference[CHI '25]{CHI Conference on Human Factors in Computing Systems}{April 26-May 1, 2025}{Yokohama, Japan}
\acmBooktitle{CHI Conference on Human Factors in Computing Systems (CHI '25), April 26-May 1, 2025, Yokohama, Japan}
\acmDOI{10.1145/3706598.3714313}
\acmISBN{979-8-4007-1394-1/25/04}

\usepackage{booktabs}
\usepackage{arydshln}
\usepackage{verbatim}
\usepackage{csquotes}
\usepackage{xspace}
\usepackage{mdwlist}
\usepackage{subfigure}
\usepackage{makecell}
\usepackage{array}
\usepackage{enumitem}
\usepackage{colortbl}
\usepackage{dsfont}
\usepackage{tabularx}
\usepackage{bbm}
\usepackage{tikz}
\usepackage{pgfplots}
\usepackage{soul} 
\usepackage{listings}
\usepackage[tableposition=top]{caption}
\usepackage{color,xcolor}
\usepackage{multirow} 
\usepackage{array}
\usepackage{longtable} 
\setlist[itemize]{left=0pt}
\newcommand{\chirev}[1]{{\color{black}{#1}}} 
\newcommand{\chifinal}[1]{{\color{black}{#1}}} 

\newcommand{\eg}{{e.g.}}
\newcommand{\ie}{{i.e.}}

\begin{document}

\title[Unlocking Scientific Concepts with LLM-generated Analogies]{Unlocking Scientific Concepts: How Effective Are LLM-Generated Analogies for Student Understanding and Classroom Practice?}

\author{Zekai Shao}
\authornotemark[1]
\email{zkshao23@m.fudan.edu.cn}
\affiliation{%
  \institution{Fudan University}
  \state{Shanghai}
  \country{China}
}

\author{Siyu Yuan}
\authornote{Zekai Shao and Siyu Yuan contributed equally to this research.}
\email{syyuan21@m.fudan.edu.cn}
\affiliation{%
  \institution{Fudan University}
  \state{Shanghai}
  \country{China}
}

\author{Lin Gao}
\email{lingao23@m.fudan.edu.cn}
\affiliation{%
  \institution{Fudan University}
  \state{Shanghai}
  \country{China}
}

\author{Yixuan He}
\email{yixuanhe24@m.fudan.edu.cn}
\affiliation{%
  \institution{Fudan University}
  \state{Shanghai}
  \country{China}
}

\author{Deqing Yang}
\authornotemark[2]
\email{yangdeqing@fudan.edu.cn}
\affiliation{%
  \institution{Fudan University}
  \state{Shanghai}
  \country{China}
}

\author{Siming Chen}
\email{simingchen@fudan.edu.cn}
\authornote{Siming Chen and Deqing Yang are the corresponding authors.}
\affiliation{%
  \institution{Fudan University}
  \state{Shanghai}
  \country{China}
}

\renewcommand{\shortauthors}{Zekai Shao, et al.}

\begin{abstract}
Teaching scientific concepts is essential but challenging, and analogies help students connect new concepts to familiar ideas.
Advancements in large language models (LLMs) enable generating analogies, yet their effectiveness in education remains underexplored.
In this paper, we first conducted a two-stage study involving high school students and teachers to assess the effectiveness of LLM-generated analogies \chirev{in biology and physics through a controlled in-class test} and \chirev{a classroom field study}.
\chirev{Test results suggested that LLM-generated analogies could enhance student understanding particularly in biology, but require teachers' guidance to prevent over-reliance and overconfidence. 
Classroom experiments suggested that teachers could refine LLM-generated analogies to their satisfaction and inspire new analogies from generated ones, encouraged by positive classroom feedback and homework performance boosts.
Based on findings, we developed and evaluated a practical system to help teachers generate and refine teaching analogies.
We discussed future directions for developing and evaluating LLM-supported teaching and learning by analogy.
}
\end{abstract}

\ccsdesc[500]{Human-centered computing}
\ccsdesc[300]{Human-centered computing~Empirical studies in HCI}
\ccsdesc[100]{Human-centered computing~User interface toolkits}
\ccsdesc[100]{Human-centered computing~Field studies}

\keywords{Analogy Generation, Large Language Models, Scientific Concept Understanding, Classroom Study}
  
\sloppy
\maketitle

\section{Introduction}
\label{sec:intro}

Analogy facilitates the comprehension of complex concepts by associating them with familiar ones~\cite{gentner_structure_1997,davies_analogy_1985,gentner_computational_2011,mitchell_abstraction_2021}. 
It plays a crucial role across various domains, particularly in education, science and problem-solving~\cite{vendetti_analogical_2015, thagard_analogy_1992, richland_analogy_2015,treagust_science_1992, gray_teaching_2021, oliva_teaching_2007}. 
It enhances cognitive processes such as creativity~\cite{holyoak_mental_1996,goel_design_1997,kao_how_2020}, aids in effective communication~\cite{casarett_can_2010,galesic_using_2013}, and facilitates the learning and understanding of complex concepts.
Analogies, such as ``water waves'' for ``light waves'', ``the solar system'' for ``atomic structure'', and ``hydraulic pump'' for ``heart'' are frequently employed in textbooks and classroom teaching to assist students in understanding scientific concepts.
Analogies can boost student understanding of concepts by opening new perspectives, making abstract ideas more relatable by connecting them to familiar situations, assisting in visualizing these concepts, sparking students’ interest and motivation, and taking into account their previous knowledge to reveal any misconceptions~\cite{duit1991role}.
 
The rapid advancement of Large Language Models (LLMs) has led researchers to employ these models in generating analogies that enhance concept comprehension~\cite{openai_gpt-4_2023}.
Unlike smaller language models (LMs), such as BERT~\cite{devlin-etal-2019-bert} and GPT-2~\cite{radford2019language}, that primarily address word-pairing analogies (\eg, ``king is to man as queen is to woman'')~\cite{mikolov_linguistic_2013,boteanu_solving_2015,gladkova_analogy-based_2016,chen_e-kar_2022}, LLMs are capable of creating more complex, free-form natural language analogies~\cite{bhavya_analogy_2022}.
In specific use cases, for example, researchers have explored using LLMs to generate biologically inspired analogies to foster scientific ideation~\cite{kang_biospark_2024} and to transform abstract data into vivid data analogies that enhance the understanding of readers~\cite{chen_beyond_2024}.

Initial research has explored the use of LLMs to generate concept-related analogies to assist students and teachers~\cite{bhavya2024analego}.
However, the effectiveness of LLM-generated analogies in educational settings remains underexplored, highlighting the need for a thorough evaluation to guide teachers and students in their practical application. 
We draw on two traditional education scenarios on evaluating human-made analogies to assess LLM-generated analogies: students solving problems using analogies without human intervention~\cite{thagard_analogy_1992,gick_analogical_1980,gick_schema_1983, brown_analogical_1989}, and teachers employing analogies in the classroom~\cite{vendetti_analogical_2015,richland_analogy_2004,oliva_teaching_2007}.
Evaluating LLM-generated analogies in the first scenario determine their effectiveness in assisting students with problem-solving by accuracy.
It also offer insights for developing LLM-assisted self-learning tools~\cite{gao2024fine, Lyu2024evaluating} that produce more beneficial educational analogies.
In the second scenario, evaluation helps to understand the needs and practical effectiveness of LLM-generated analogies in real classroom practices with teachers' instruction.
\chirev{It also provides insights of needs for developing LLM-assisted tools to help teachers teach with analogies.
Therefore, we aim first to evaluate the effectiveness of LLM-generated educational analogies in helping students grasp scientific concepts across two distinct educational settings. 
Based on the findings, we then consider developing a practical system to explore the practical application of LLMs for analogy generation in supporting teachers and students.
}

\chirev{Evaluating LLM-generated analogies in practical applications is challenging, requiring manual annotation~\cite{sultan_life_2022} and human-subject study involving diverse participants across varied settings.}
Previous studies have evaluated the use of LLM-generated analogies in creative tasks, such as design problem reformulation~\cite{ding_fluid_2023}. 
However, due to the varied participants and cognitive demands of education, the methodologies and findings from these domains are not transferable to educational settings.
To address the gap, we answered the following research questions (\textbf{RQ}s): 
\begin{itemize}
    \item \textbf{RQ1:} How effective are LLM-generated analogies for student understanding without human intervention? 
    \item \textbf{RQ2}: What kind of analogies do teachers need LLMs to generate for classroom practice?
    \item \textbf{RQ3}: How effective are these LLM-generated analogies for classroom practice with teacher intervention?
\end{itemize}

To answer these \textbf{RQ}s, we design a two-stage study to evaluate the effectiveness of LLM-generated analogies in helping students understand scientific concepts in educational settings.
We use GPT-4o~\cite{openai_gpt-4_2023}, the state-of-the-art LLM, to generate analogies for concepts in physics and biology, and conduct human-subject study in a \chifinal{Chinese} high school.
We addressed \textbf{RQ1} through a controlled in-class students test in \textbf{Study I}: the experimental group used both textbook explanations and carefully generated analogies to answer questions about unknown concepts, while the control group relied only on textbook explanations.
The effectiveness of LLM-generated analogies was primarily measured by accuracy, with self-confidence rating evaluating students' subjective satisfaction. 
Interviews with participating students provided additional insights.
In \textbf{Study II}, we conducted pre-class interviews with teachers and senior students to understand the need for LLM-generated analogies in classroom teaching (\textbf{RQ2}). 
Interview findings helped us design the following field study and refine analogy generation.
We then carried out a one-week field study with twelve lessons in which teachers selected and modified the generated analogies based on their needs, using them in one class while maintaining regular teaching in the other.
Through observations and teacher interviews, we qualitatively evaluated the effectiveness of LLM-generated analogies for classroom teaching with teacher intervention (\textbf{RQ3}).

\chirev{Based on the findings from both studies, we designed an interactive system that uses LLM to assist teachers in preparing analogies. 
We first interviewed two teachers to assess what they want from the system. 
Using insights from the interviews and studies, we defined the design requirements to guide system development.
We invited six physics and biology teachers from different schools to participate in the system evaluation. 
After tutorial and free exploration to confirm their familiarity with the system, teachers used the system to create analogy of scientific concepts for teaching over a week. 
Based on users' data records and interviews, we demonstrated the practical effectiveness of LLMs in supporting teaching by analogy and then discussed future research directions.
}

The main \textbf{contributions} are summarized as follows:
\begin{itemize}
    \item We are among the first to design and conduct a two-stage study, comprising a controlled experiment and a field study, to evaluate the effectiveness of LLM-generated analogies in student understanding and classroom practice.
    \item We contribute \chirev{empirical evidence and} new knowledge into educational LLM-generated analogies, revealing that their effectiveness without intervention \chirev{varies} on subject characteristics and can lead to student overconfidence, while in classroom practice, analogies are refined by teachers to align with their teaching focus and preference, enhancing both classroom and homework performance and inspiring new teaching methods.
    \item \chirev{Based on empirical evidence and new knowledge, we developed a practical system to help teachers build analogies of scientific concepts, conducted a system evaluation demonstrating its effectiveness, and provided design implications for future development and evaluation of LLM-assisted education with analogy.}
\end{itemize}

\section{Related Work}
\label{sec:related}
This section reviews related work on analogy in education, evaluating analogy in HCI, analogy-making with LLMs, and \chirev{LLM-assisted educational systems.} 
\subsection{Analogy in Education}
Analogies help humans understand complex concepts by linking them to familiar ones, making them a valuable tool in educational contexts. 
Many studies~\cite{thagard_analogy_1992,gick_analogical_1980,gick_schema_1983, brown_analogical_1989} have investigated analogical problem solving, where students of various ages solve unfamiliar problems using well-designed analogies. 
Through observational feedback and statistical analysis, researchers have established frameworks and several guidelines for using analogies in education. 
For example, as discussed by~\cite{gick_analogical_1980}, the source of the analogy would share similar relationships with the target, yet originate from a semantically distant field. 
However, such lab studies often involve experimenters posing problems, with students merely solving them without instructional guidance~\cite{brown_analogical_1989}, which diverges from real classroom learning.
Therefore, further research~\cite{vendetti_analogical_2015,richland_analogy_2004,treagust_science_1992,oliva_teaching_2007} have investigated how teachers and students engage with analogies in classroom settings, leading to nuanced insights on the influence of students' age and background and teachers' strategies.

Although previous studies have explored the characteristics and use of analogies in education, they have not examined those generated by LLMs, which is crucial given the growing importance of LLM-assisted education~\cite{gao2024fine, Lyu2024evaluating}.
Our work fills this gap by leveraging LLMs to generate analogies tailored to specific education needs, incorporating established characteristics from prior literature and our interviews.
We design human-subjective studies to evaluate their effectiveness in problem-solving tests and classroom environments following prior research.

\subsection{Evaluating Analogy in Human-Computer Interaction}
Analogy has long been studied in HCI for its effectiveness in various context, including algorithms improvement~\cite{bureaucracy2020Pääkkönen,streetlevel2019Alkhatib}, cancer communication~\cite{capturing2024hnatyshyn}, narrative framing~\cite{reelframer2024wang}, enhancing deliberation~\cite{help2024yeo}, communicating standardized effect sizes~\cite{putting2022kim}, and sensemaking of LLM responses~\cite{supporting2024gero}.

Two key research directions about analogies in HCI are for enhancing numerical comprehension through data analogy and fostering creativity.
Data analogies link abstract data to familiar concepts to improve understanding. Researchers evaluate these analogies using controlled experiments and assess effectiveness through subjective ratings like helpfulness~\cite{toput2018riederer, improving2018hullman, generating2016kim, spatharioti_using_2024, chen_beyond_2024}, estimation errors~\cite{toput2018riederer, improving2018hullman}, and correlations between model and human ratings~\cite{spatharioti_using_2024}.
Analogies also facilitate scientific discovery and design. 
In scientific discovery, evaluations involve coding analogy types~\cite{solvent2018chan, kang_augmenting_2022}, calculating similarity metrics~\cite{solvent2018chan}, and conducting think-aloud sessions with scientists~\cite{kang_augmenting_2022}. 
For creative design, analogies are assessed by novelty~\cite{searching2014yu, bilogically2023zhu, bidtrainer2024chen}, quality~\cite{distributed2014yu, bidtrainer2024chen}, relevance and domain distance~\cite{analogymining2018Gilon}, feasibility~\cite{bilogically2023zhu}, and rationality~\cite{bidtrainer2024chen}.
Recently, Ding et al.~\cite{ding_fluid_2023} explored GPT-3's capacity to augment cross-domain analogical reasoning, finding it helpful for creative problem reformulation despite the risks of harmful content.

However, there has been limited exploration of analogy search in HCI for education~\cite{kumar2015stickipedia}. 
While researchers have adopted LLMs to help students and teachers generate novel analogies~\cite{bhavya2024analego}, systematic evaluations of their effectiveness in educational settings are lacking. 
Given the unique cognitive demands of education, existing assessments~\cite{ding_fluid_2023} may not be directly applicable. 
Our work aims to address this gap and offer insights into analogy generation for education.

\subsection{Analogy-making with Language Models}
Analogy is vital for human cognition and has attracted considerable interest from the AI research community. 
Traditionally, studies on analogy-making in AI have concentrated on creating word analogies (\eg, ``king is to man as queen is to woman'') using smaller language models (LMs), e.g., BERT~\cite{devlin_bert_2019} and GPT-2~\cite{radford_language_2019} trained on specific datasets~\cite{turney_combining_2003,mikolov_linguistic_2013,boteanu_solving_2015,gladkova_analogy-based_2016,chen_e-kar_2022, yuan_analogykb_2023}.
With the advancement of LLMs~\cite{ouyang_training_2022,team_gemini_2023,touvron_llama_2023,openai_gpt-4_2023}, there has been a shift toward generating natural language analogies, \ie, free-form analogies~\cite{bhavya_analogy_2022,webb_emergent_2022,ding_fluid_2023,wijesiriwardene_analogical_2023,jiayang_storyanalogy_2023,hu_-context_2023,sultan_parallelparc_2024} and forming structural analogies~\cite{sultan_life_2022,yuan_beneath_2023}.
Researchers typically design prompts manually for free-form analogies to guide LLMs in \chifinal{generating analogies~\cite{bhavya_analogy_2022, webb_emergent_2022}.
For example, Bhavya et al.~\cite{bhavya_analogy_2022} constructed a new dataset including standard science analogies and science analogies from academics and adopted prompt engineering to ask LLMs to generate analogies. 
The results show that LLMs are sensitive to prompt design, temperature, and injected spelling errors, particularly the distinction between questions and imperative statements.
We followed their optimal prompt format for our generation process.
}
\chirev{
For evaluation of the generation quality of analogy, previous studies have relied on annotators manually evaluating analogies according to established principles of analogy cognition~\cite{sultan_life_2022}. 
}

In contrast to these approaches, our study is pioneering in investigating how analogies generated by LLMs can help students understand scientific concepts. 
We analyze the characteristics of analogies in educational settings through literature reviews and interviews and incorporate them into prompts for generation. 
Then, we use LLMs to generate educational analogies and evaluate them in real tests, classroom practice, \chirev{and a practical system}.

\subsection{\chirev{LLM-assisted Educational Systems}}
\chirev{
With the rapid advancement of LLMs, researchers are exploring their potential to develop efficient and practical systems that support students and teachers in educational tasks~\cite{kasneci2023chatgpt,yan2024practical}. 
For students, many studies have focused on creating intelligent tutoring systems powered by LLMs. 
Examples include enabling fully autonomous self-learning pipelines to support self-regulated learning~\cite{gao2024fine} and developing and evaluating LLM-based learning assistants in classroom settings~\cite{kazemitabaar_codeaid_2024,Lyu2024evaluating}.
For teachers, several LLM-based systems are designed to effectively monitor and analyze students' learning activities~\cite{ngoon_classinsight_2024,cflow,vizgroup}.
In addition, researchers aim to assist teachers in creating diverse teaching materials, such as lesson plans~\cite{lessonplanner2024uist}, diagrammatic problems~\cite{edgeworth}, and reading quizzes~\cite{readingquizmaker}.

Our work explores a novel aspect of LLM-driven education: evaluating the effectiveness of LLMs in generating teaching analogies. 
One preliminary research has initially explored generating educational analogies~\cite{bhavya2024analego}, while its system design lacks the support of empirical evidences and fails to address teachers' needs.
Instead, we first conducted a two-stage study to gain insights and empirical evidence and identify needs for teachers and students. 
We then developed and tested a system to support teachers in creating and refining analogies for lesson preparation and discussed future integration with diverse LLM-based educational tools for various users.

}

\section{Method}
\label{sec:method}
In this section, we introduce the overview, our study design, and the techniques for analogy generation with LLM.

\begin{figure*}[h]
    \centering
    \includegraphics[width=0.9\linewidth]{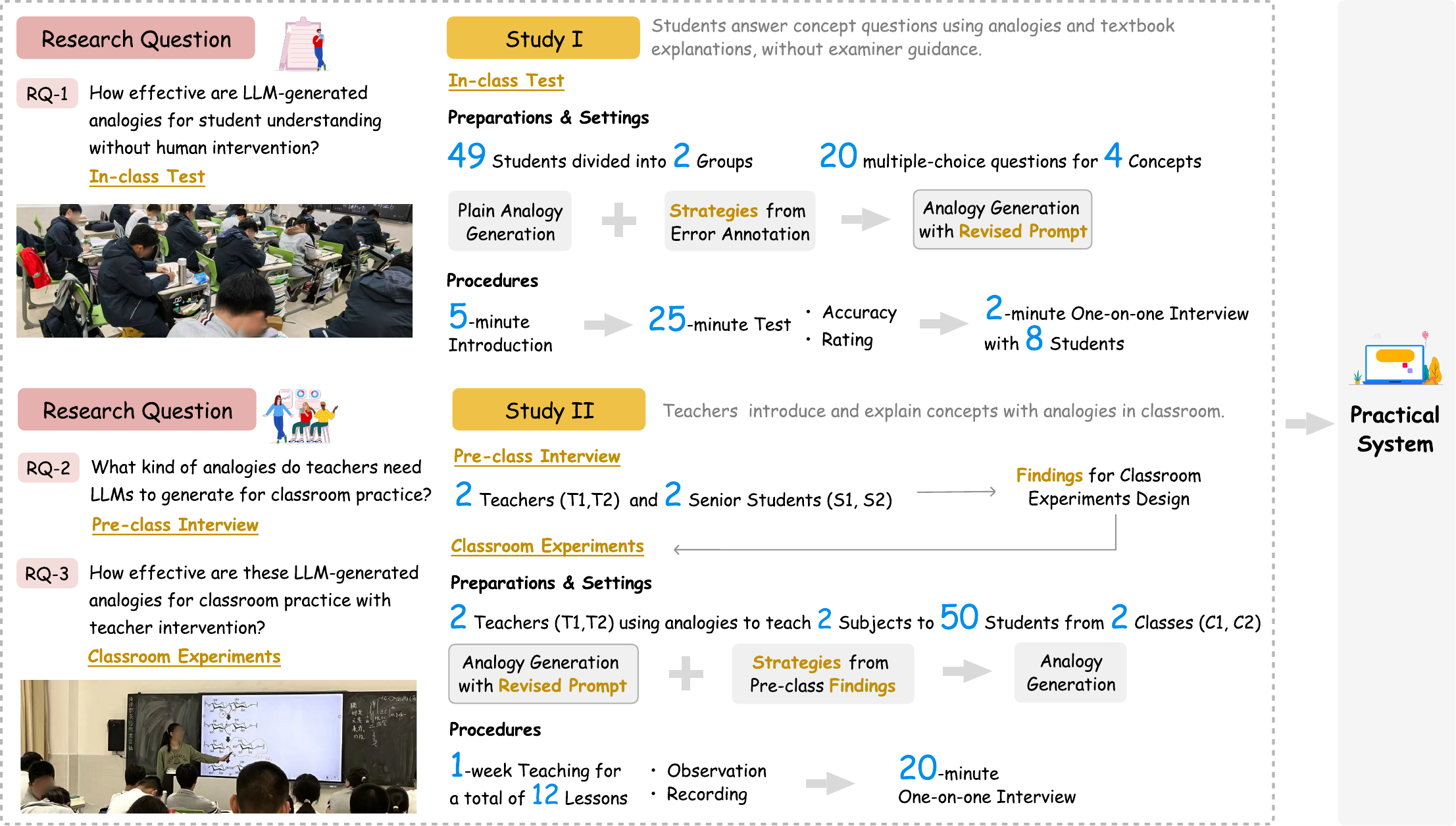}
    \caption{\chirev{Our Study I address RQ1 through an in-class test. Study II address RQ2 through a pre-class interview and RQ3 through a one-week classroom study. Studies' findings lead us to build an LLM-supported practical system for analogical education.}}
    \Description{The figure presents our two-stage study design addressing three research questions (RQ1, RQ2, and RQ3), which then direct us to design and evaluate an LLM-supported practical system for education by analogy. Study I investigates the effectiveness of large language model-generated analogies for student understanding, assessed through an in-class test and student interviews. Study II explores the types of analogies teachers require and their effectiveness in classroom practice through pre-class interviews and a week-long field study with classroom teaching. }
    \label{fig:overview}
\end{figure*}

\chirev{
\subsection{Overview}
Our work aims to first understand LLM-generated educational analogies' effectiveness through empirical studies and then design practical LLM-assisted educational systems leveraging findings from studies.
For empirical studies, we explore two study settings: one where students solve problems using only LLM-generated analogies and necessary materials without additional guidance (Sec.~\ref{sec:study1}), and another where teachers integrate LLM-generated analogies flexibly into classroom instruction (Sec.~\ref{sec:study2}), considering the following two reasons.
First, these two settings align with those used to evaluate human-made analogies in traditional education research: student-only testing~\cite{thagard_analogy_1992,gick_analogical_1980} and teacher-led classroom practice~\cite{vendetti_analogical_2015,oliva_teaching_2007}.
Second, evaluating in two settings respectively inform the design of systems that incorporate LLM in generating analogies to (1) support self-learning for students~\cite{gao2024fine} and (2) boost teaching for teachers~\cite{lessonplanner2024uist}, within the context of LLM-assisted education research.

Based on study findings, we identify a more feasible direction between supporting student self-learning and teacher instruction by analogy for designing a practical system (Sec.~\ref{sec:system}). 
Drawing from study findings and system evaluation, we offer implications for future LLM-assisted educational systems with analogy (Sec.~\ref{sec:discussion}).

}

\subsection{Study Design}

\chirev{
For both studies, we need to ground the effectiveness of LLM-generated analogies on improving students' concept understanding by comparing the accuracy of problem-solving across different student groups.
Specifically, we may need to compare students' problem-solving accuracy in classroom teaching with and without LLM-generated analogies. 
This comparison is driven by the unique features of LLM-generated analogies compared to traditional ones, as well as teachers’ potential unfamiliarity with them and unclear expectations.
Unlike classic analogies refined and validated over generations, they may be harder for teachers to adapt to support student understanding.
We will confirm this design in the interview with teachers (Sec.~\ref{sec:study21_findings_and_derived_requirments}).
} 
Besides accuracy, it's important to evaluate students' and teachers' subjective satisfaction~\cite{Lyu2024evaluating}, reflected through subjective ratings, classroom feedback, and interviews.

As shown in Fig.~\ref{fig:overview}, we conducted a two-stage study in a Chinese high school on LLM-generated analogies for physics and biology concepts to explore this topic. 

\chifinal{
\subsubsection{Study I} 
We conducted an in-class test to evaluate the effectiveness of LLM-generated analogies in understanding concepts without human intervention (\textbf{RQ1}).}
\begin{itemize}
    \item \textbf{Participants.} 49 Chinese high school freshmen from 2 classes.
    \item \textbf{Procedure.} Students were divided into two groups: one received LLM-generated analogies, while the other did not. They then completed an in-class test with 20 multiple-choice questions for 4 concepts they didn't learn. Afterward, 8 students participated in interviews.
    \item \textbf{Measure.} Effectiveness was assessed through quantitative results of students' answer accuracy and confidence ratings, and qualitative insights from student interviews.
\end{itemize}

\chifinal{
\subsubsection{Study II} Study II consists of two sub-studies.}
We conducted a pre-class interview to identify classroom needs for LLM-generated analogies (\textbf{RQ2}).
\begin{itemize}
    \item \textbf{Participants.} 2 Chinese teachers and 2 Chinese senior students.
    \item \textbf{Procedure.} The interview followed a semi-structured format, allowing participants to discuss their experiences, expectations, and concerns on using LLM-generated analogies in the classroom.
    \item \textbf{Measure.} We identified qualitative findings on their perceptions of analogy use during the interview.
\end{itemize}

We then conducted a controlled field study to evaluate the effectiveness of LLM-generated analogies in classroom practice (\textbf{RQ3}).
\begin{itemize}
    \item \textbf{Participants.} The 2 teachers from the pre-class interview and 50 Chinese students from 2 classes.
    \item \textbf{Procedure.} The teachers taught both classes over one week, delivering a total of 12 lessons. In one class, they incorporated LLM-generated analogies, while in the other, they followed regular instruction without analogies as a control.
    \item \textbf{Measure.} We derived qualitative findings from teachers’ selection and modification of analogies and student feedback.
\end{itemize}



\subsection{Techniques for Analogy Generation with LLM}
\label{sec:Techniques for Analogy Generation with LLM}


With the development of LLMs, their capabilities have demonstrated significant potential in generating satisfying content.
Therefore, recent studies have utilized LLMs to create analogies using manually-designed instructions. 
We align with this approach by crafting prompts for LLMs grounded in established analogy theories and our interviews.
The prompt consists of three parts:
\begin{itemize}
    \item \textbf{Task Description} demonstrates the task that LLMs need to achieve.
    \item \textbf{Principles} highlight the requirements, rules, and constraints that LLMs must follow to complete the task.
    \item \textbf{Input Resource} lists the input materials needed to complete the task.
\end{itemize}

\chirev{
Given the lack of standardized guidelines for using LLMs in educational analogy generation, we summarize principles from educational literature~\cite{hesse1959defining,gentner1983structure,gentner2017analogy} and refine them through manual annotation in \textbf{Study I} and interviews in \textbf{Study II}. 
Additionally, inspired by prior work~\cite{yuan-etal-2023-distilling}, Study I uses an over-generation and filtering strategy to select the best analogies, while Study II leaves all candidates for teachers to refine.
The techniques used in the further developed system is determined by the results of two studies.
}


\section{Study I}
\label{sec:study1}
In this section, we detail the participants, data preparation process, stimuli, procedure, and results analysis of Study I.


\subsection{Participants} 
\label{sec:study1_participants}
Two classes of freshmen, totaling 49 \chifinal{Chinese} students from a Chinese high school with which we have a scientific research collaboration, participated in Study I. 
Their ages ranged from 15 to 17, with 26 males and 23 females. 
They had recently started high school physics and biology courses. 
Their entrance exam scores and classroom performance suggested a normal cognitive level, and we did not pre-select students based on their abilities.

\subsection{Data Preparation}
\label{sec:study1_data_preparation}

\begin{figure*}[tb]
    \centering
    \includegraphics[width=0.9\linewidth]{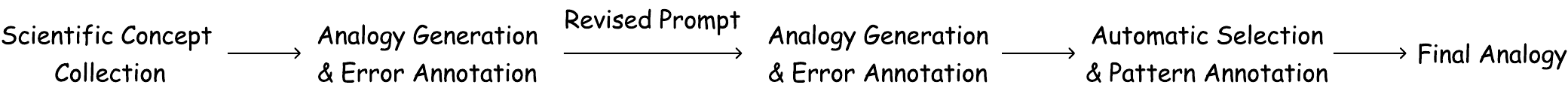}
    \caption{The pipeline for analogy generation to explain scientific concepts for data preparation in Study I.}
    \Description{This figure presents our pipeline for generating and refining analogies to explain scientific concepts, as used in Study I for data preparation. The process begins with collecting scientific concepts and generating initial analogies using LLMs. Errors are then annotated, and the analogies are refined with revised prompts. Next, the most effective analogy is selected automatically, and patterns are annotated for the final selection.}
    \label{fig:pipeline}
\end{figure*}
\begin{table*}[t]
\footnotesize
  \centering
  \caption{\chifinal{Prompt template for GPT-4 to generate analogies (I) and select the analogy from three candidates for the given concept (II).
  {\color[rgb]{0,0.39,0}{\textit{Green texts}}} are new principle after revising.}
  }
    \begin{tabularx}{0.9\linewidth}{X}
    \toprule
    
    \multicolumn{1}{c}{\cellcolor[gray]{0.95}{\textbf{I: Analogy Generation with Revised Prompt}}} \\
    \midrule
    \makecell[l]{\color{gray}{/* \textit{Task Description} */}\\
    Your task is to use an analogy to explain the scientific concept to students. \\
    Here are some principles for generating appropriate analogies: \\
    \color{gray}{/* \textit{Principles} */}\\
    1. The similarity between the objects in the analogy and those in the scientific concept should be minimal.\\
    2. The relationships in the analogy and the scientific concept should be highly similar.\\
    3. The analogy should use objects that students are very familiar with from everyday experiences.\\
    \color[rgb]{0,0.39,0}{\textit{4. The analogy should accurately identify similar relationships with the scientific concept and avoid forcing non-existent similarities.}}\\
    \color[rgb]{0,0.39,0}{\textit{5. The objects in the analogy and the scientific concept should align with scientific laws and commonsense knowledge.}}\\
    \color[rgb]{0,0.39,0}{\textit{6. An object in the analogy cannot have different roles or functions in different contexts.}}\\
    \color[rgb]{0,0.39,0}{\textit{7. The logic within a sentence or paragraph should not be self-contradictory.}}\\
    \color{gray}{/* \textit{Input Resource} */}\\
    To generate the appropriate analogy according to students' learning progress, we provide you with textbook content related to this scientific concept\\ for your reference.\\
    The textbook content: \texttt{\{input\_text\}}\\
    The scientific concept: \texttt{\{concept\}}\\
    Analogy:
     }\\
     \midrule
\multicolumn{1}{c}{\cellcolor[gray]{0.95}{\textbf{II: Analogy Selection}}} \\
    \midrule
    \makecell[l]{\color{gray}{/* \textit{Task Description} */}\\
    There are three candidate analogies which are used to explain the scientific concept based on textbook content. Your task is to select the best analogy\\ from these three candidates. \\
    Here are some principles for generating appropriate analogies: \\
    \color{gray}{/* \textit{Principles} */}\\
    \color{gray}{Same as principles in I (Omit)}\\
    \color{gray}{/* \textit{Input Resource} */}\\
    The textbook content: \texttt{\{input\_text\}}\\
    The scientific concept: \texttt{\{concept\}}\\
    The generated analogies: \\
    Candidate 1: \texttt{\{analogy\_1\}}\\
    Candidate 2: \texttt{\{analogy\_2\}}\\
    Candidate 3: \texttt{\{analogy\_3\}}\\
    You need to give reasons first and then give the answer with the format: \texttt{Final Answer: Candidate X} \\
    Answer:
     }\\
    \bottomrule
    \end{tabularx}
  \label{tab:instruction_prompt}
\end{table*}

\begin{table*}[t]
\centering
\small
\caption{Errors and accuracy of Plain Generation (\textbf{Plain}), Revised Generation (\textbf{Revised}), and Automatic selection (\textbf{Selection$_\texttt{Auto}$}). (\textbf{Data \#}) shows the number of data. \chifinal{($\downarrow$) indicates lower values are better, while ($\uparrow$) indicates higher values are better.}}
\label{tab:error_rates}
\begin{tabular}{lccccccc}
\toprule
\multirow{3}{*}{\textbf{Process}}  & \multirow{3}{*}{\textbf{Data \#}} & \multicolumn{2}{c}{\textbf{Factuality}}     & \multicolumn{2}{c}{\textbf{Consistency}}        & \multirow{3}{*}{\textbf{Accuracy} ($\uparrow$)} \\
\cmidrule(lr){3-4} 
\cmidrule(lr){5-6} 
                      &  & \textbf{Analogy Object}& \textbf{Inappropriate} & \textbf{Object} & \textbf{Logical} & \\ 
                      & & \textbf{Paradox}  ($\downarrow$)& \textbf{Analogy} ($\downarrow$)& \textbf{Confusion} ($\downarrow$)& \textbf{Contradiction} ($\downarrow$)& \\
\midrule
Plain         &     30  & \textbf{0.27}                & 0.23                  & 0.23                & 0.03                 & 0.53    \\
Revised      &      30     & 0.33           & 0.23                  & \textbf{0.17}                & \textbf{0.00}                 & 0.53    \\
Selection$_\texttt{Auto}$   &      10     & 0.30           & \textbf{0.20}                  & 0.20              & \textbf{0.00}                 & \textbf{0.60}    \\ 
\bottomrule
\end{tabular}
\end{table*}
\chirev{As shown in Fig.~\ref{fig:pipeline}, we began by manually selecting ten scientific concepts from physics and biology in Chinese high school textbooks. 
Next, we used the advanced LLM, GPT-4o~\cite{openai_gpt-4_2023} (temperature = 0.7) to generate three analogies for each concept with three principles summarized from education research~\cite{hesse1959defining,gentner1983structure,gentner2017analogy} incorporated in the prompt. 
Three authors independently identified and annotated errors in generated analogies. 
After repeated discussion, the annotators classified the errors into four types: two related to factuality and two to consistency, as follows.
\begin{itemize}
    \item \textbf{Analogy Object Paradox}: The objects of the analogy do not align with physical laws or commonsense knowledge.
    \item \textbf{Inappropriate Analogy}: The analogy fails to accurately mirror the concept, leading to misconceptions.
    \item \textbf{Object Confusion}: The same analogy objects are assigned different roles or functions across various contexts.
    \item \textbf{Logical Contradiction}: The syntax within a sentence or paragraph contradicts itself.
\end{itemize}

The inter-rater reliability among annotators reached Fleiss' Kappa of 0.83 for Analogy Object Paradox, 0.94 for Object Confusion, and 1 for the remaining error codes.
Error annotations in subsequent steps achieved similar reliability.
As shown in the first row of Tab.~\ref{tab:error_rates}, out of the 30 generated analogies, 16 were correct. 
The remaining analogies frequently exhibited the first three error types with one analogy containing logical contradiction.
From these errors, we derived four new principles and added them to the prompt template (Tab.~\ref{tab:instruction_prompt} I) to help GPT-4o avoid these errors.
However, even with these improvements, GPT-4o still made errors. 
To address this, we followed prior AI research~\cite{pan2023automatically,yuan-etal-2023-distilling,liu2024large} and allowed GPT-4o to automatically select the best of the three candidate analogies generated for each concept.
\chifinal{The prompt for analogy selection is shown in Tab.~\ref{tab:instruction_prompt} II.}
As shown in the third row of Tab.~\ref{tab:error_rates}, enabling the model to self-correct improved the accuracy of the analogies.}

\begin{figure*}[t]
    \centering
    \includegraphics[width=0.9\linewidth]{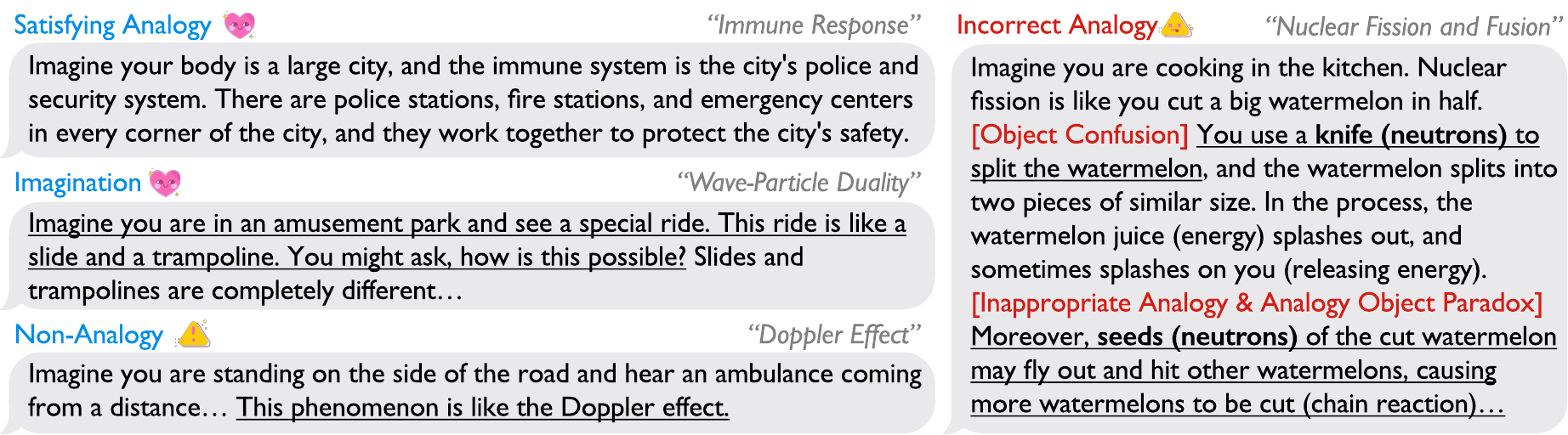}
    \caption{\chirev{The examples of final analogies generated from GPT-4o after iterative generation and annotation in Study I.}}
    \Description{This figure shows examples of analogies generated by GPT-4o after multiple iterations. In the ``Correct'' Analogy category, a satisfying analogy for ``immune response" compares the body to a city protected by a security system. An imaginative analogy for ``wave-particle duality" likens it to an amusement park ride, though less accurate. A non-analogy for the ``Doppler effect" fails to form a true analogy. In the Incorrect Analogy category, the analogy for ``nuclear fission'' confuses objects by comparing neutrons to a knife and seeds, leading to inappropriate and paradoxical explanations.}
    \label{fig:example}
\end{figure*}


Finally, we further categorized ten analogies selected by the LLM into four distinct groups, as illustrated in Fig.~\ref{fig:example}:
\begin{itemize}
    \item \textbf{Correct and Satisfying Analogy}: Analogies in this category are error-free. 
    The objects in these analogies are realistic, align with common sense, and adhere to physical laws, effectively and vividly illustrating scientific concepts.
    \item \textbf{Correct Analogy with Imagination}: Analogies in this category require envisioning non-existent objects or processes to explain a concept. While logically sound, they demand creative thinking and imagination from students.
    \item \textbf{Correct Non-Analogy}: This is more akin to an example-based explanation than a true analogy and is not generally recognized as an analogy in cognitive science.
    \item \textbf{Incorrect Analogy}: This category includes analogies exhibiting previously identified error types. 
    These analogies are inappropriate for students to refer to, as they do not accurately convey the intended concept.
\end{itemize}

The analogies for the five biological concepts fell under the \textbf{Correct and Satisfying Analogy} category. 
In contrast, the five physical concepts were distributed as follows: three under \textbf{Incorrect Analogy}, one under \textbf{Correct Analogy with Imagination}, and one under \textbf{Correct Non-Analogy}.

We limited the number of concepts and analogies to four to avoid overwhelming students with too much new knowledge in further tests. 
To ensure a balance across subjects and categories, we specifically selected two biological concepts categorized as \textbf{Correct and Satisfying Analogy} and two physical concepts, one each under  \textbf{Correct Analogy with Imagination} and \textbf{Incorrect Analogy}, as shown in the right side of Fig.~\ref{fig:example}.
We excluded the one \textbf{Correct Non-Analogy} from further consideration, as it is not typically classified as an analogy.

\subsection{Stimuli}
\label{sec:study1_stimuli}
Since students do not have access to electronic devices and are more familiar with and serious about traditional classroom tests, we conducted offline tests in class.

Based on the data preparation, we printed a test paper and two reference materials. 
The test paper comprises 20 multiple-choice questions, with 5 questions assigned to each of the following 4 concepts: Nuclear Fission and Fusion, Wave-Particle Duality, Blood Sugar Regulation, and Immune Response. 
In addition to selecting answers, students are required to complete a 5-point Likert scale rating for self-confidence to measure their subjective satisfaction.
The first reference material provides textbook explanations for the four concepts, while the second adds LLM-generated analogies before the explanations.
The test paper and the reference materials present the concepts in the same order. 
They are highlighted in bold, making it easier for students to find and connect the information with the questions.

\subsection{Procedure}
Then, we conducted an in-class test for students in two classes lasting 30 minutes. 
We first gave a 5-minute introduction for the background of our test.
After the introduction, we randomly divided the students into two groups and distributed two sets of reference materials to each group. 
We clarified the meaning of self-confidence rating. 
Under our supervision, each student then independently completed the test using the materials provided in 25 minutes.

After the test, we interviewed four students from each group, totaling eight participants.
Each 2-minute interview earned participants a \$2 gift card.
We asked them about any difficulties during the test and, for those with analogies in their materials, how these helped them answer questions alongside textbook concepts.

\subsection{Results Analysis}
\label{sec:study1_results_analysis}
Our selection criteria excluded test papers with incomplete answers. 
After examining the 49 test papers, we excluded 5 that had more than 5 unanswered questions. 
The remaining 44 fully completed papers, 22 from each group, were considered valid data.
Our analysis followed a top-down approach, starting from the overall test (20 questions) and proceeding to finer levels: subject (10 questions each), concept (5 questions each), and individual questions. For further comparison, students' responses were averaged across questions at the first three levels.


\chirev{
We computed descriptive statistics to gain overall insights and performed statistical tests to determine significance at each level. 
The experiment results include the students' answer accuracy and confidence ratings, both can be regarded as ordinal categorical variables.
Thus, we mostly employed the exact Wilcoxon-Mann-Whitney test (using the R package \texttt{coin}) to evaluate the significance of difference between the two groups. 
For one exception, we employed Fisher's exact test (using the R package \texttt{stats}) on students' answer accuracy at the individual question level, where the accuracy is binary (either 0 or 1). 
In any of the tests, a small p-value indicates a potential association between the use of LLM generated analogies and the students’ outcomes, and a significance level is defined as $p$<0.05 in all tests.
We also calculated Kendall's tau correlation coefficient to assess the relationship between students’ objective answer accuracy and subjective confidence ratings within each concept and group.
}

We summarize our findings as follows. The two groups are referred to as Group T (Textbook explanation only) and Group L (Textbook explanation with LLM-generated analogy), while the interviewed students are denoted as T1-T4 and L1-L4.

\begin{figure}[t]
    \centering
    \includegraphics[width=1\linewidth]{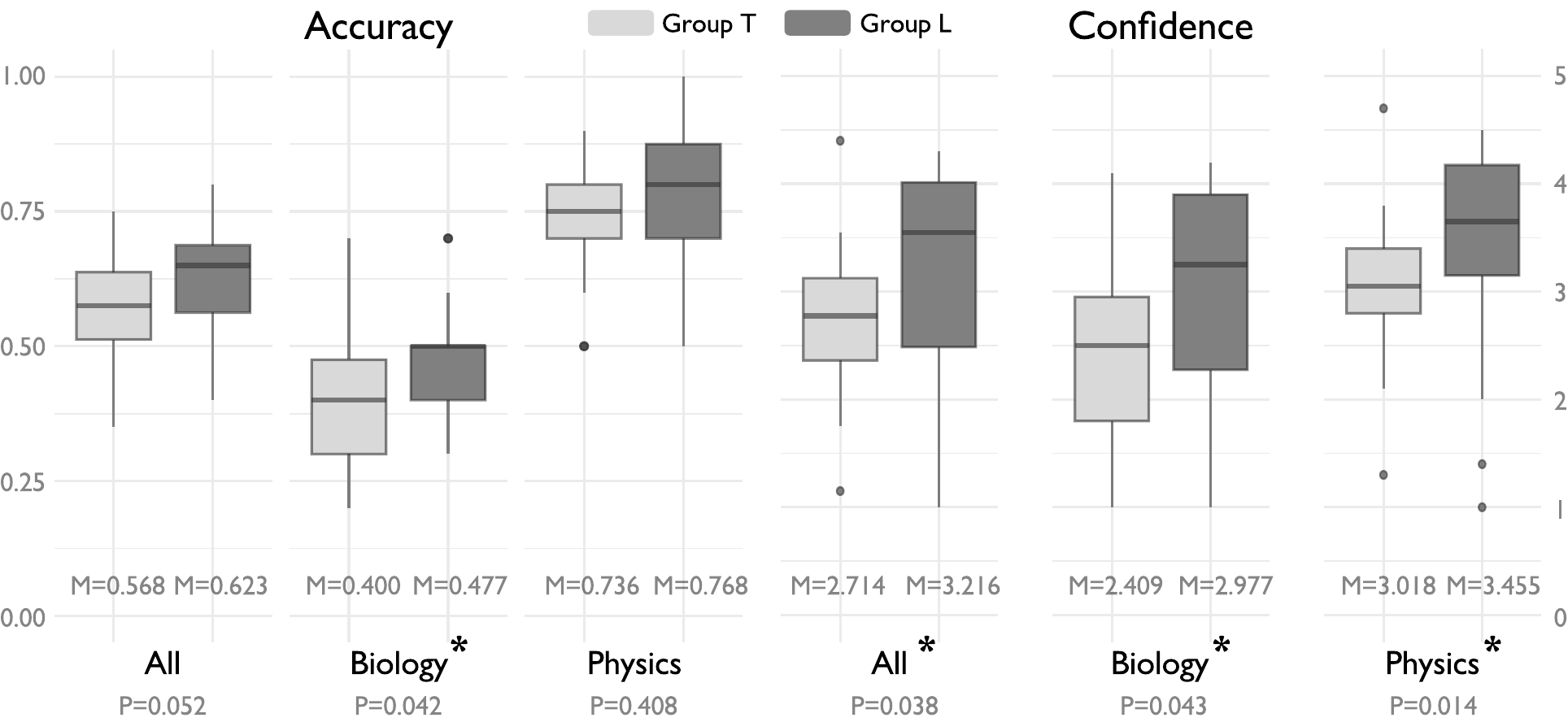}
    \caption{Boxplots showing the distribution of answer accuracy and confidence ratings for the two groups and comparing the overall test results and the two subjects. ``M'' represents Mean, ``p'' represents the significance of the association between accuracy and group, determined by the exact Wilcoxon-Mann-Whitney test, and * represents significance ($p$<0.05).}
    \label{fig:study1_acc_conf_overall}
    \Description{This figure includes boxplots showing the distribution of answer accuracy and confidence ratings. The plot has facets of accuracy and confidence rating. In each facet, X-labels are all questions, biological questions, and physical questions. The two boxplots of two groups are juxtaposed inside each label. In the accuracy facet on the left, all boxplots share the same 0 to 1 y-axis. Physics questions have overall higher accuracy with no significant difference between groups. Biology questions have overall lower accuracy with significant differences between groups. All questions are thus in the midst with no significant difference between groups. In the confidence rating facet on the right, all boxplots share the same 0 to 5 y-axis. Physics questions have a mildly higher confidence rating at each quantile while biology questions are mildly lower, but the boxplots from different labels largely overlap. All differences in confidence ratings between groups are significant.}
\end{figure}

\begin{figure}[t]
    \centering
    \includegraphics[width=1\linewidth]{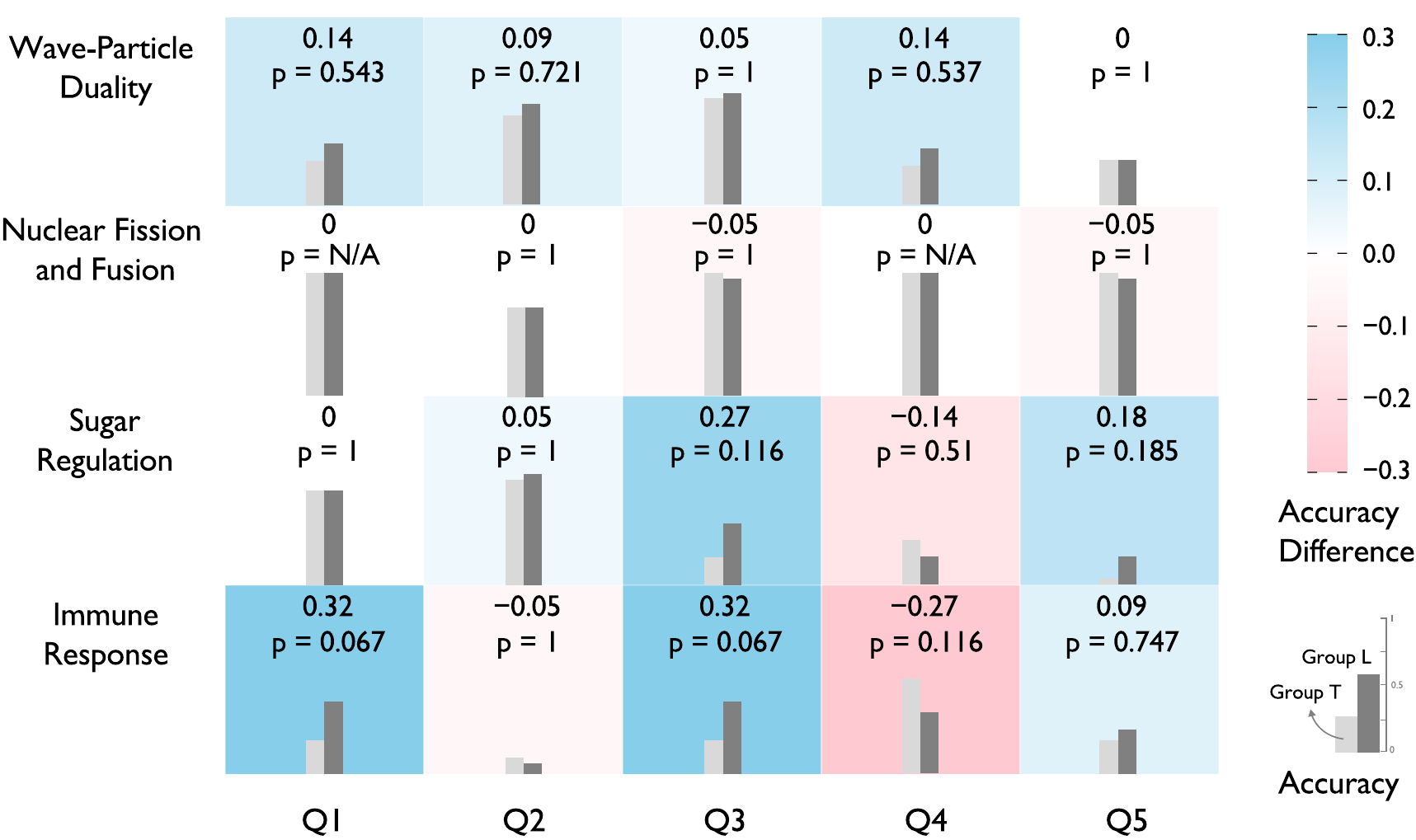}
    \caption{Heatmap of accuracy differences between Group L and Group T for individual questions, with blue indicating higher accuracy for Group L and red for Group T. Each cell contains a bar chart of the respective accuracies and a p-value representing the significance of the association between accuracy and group, determined by Fisher's exact test. For two of the questions, since all students in both groups answered them correctly, Fisher's exact test is not applicable, and $p$ = N/A.}
    \label{fig:study1_acc_question}
    \Description{This figure includes a Heatmap of accuracy differences between Group L and Group T for individual questions. The cells of heatmap are vertically divided by concepts, and then horizontally divided by the question ID in each concept. For ``Immune Response'' and ``Sugar regulation concept'', student accuracy shows much more significant differences in part of the questions (Q1, Q3, Q4 of ``Immune Response'' and Q3 of ``Sugar regulation concept''). For most questions, Group L had higher accuracy, except for Q3 ``Immune Response'' in which Group L showed clearly lower accuracy. However, for Wave-Particle Duality and Nuclear Fission and Fusion concept, the comparison between student accuracy from each group is ambiguous, as all the superiority or inferiority in accuracy is negligible considering the p-value. }
\vspace*{-10pt}
\end{figure}

\textbf{LLM-generated analogies generally aid problem-solving and have a greater impact on biological concepts than physical concepts.}
As shown in the left of Fig.~\ref{fig:study1_acc_conf_overall}, the overall accuracy for physics questions is higher, while a significant association exists between accuracy and group for biology questions ($p$ = 0.042).
Examining individual questions (Fig.~\ref{fig:study1_acc_question}), there are three questions within the two biological concepts where Group L's accuracy largely exceeds Group T's by more than 0.25.
Besides, a strict Fisher’s exact test shows marginally significant associations between accuracy and group for two questions (Q1 and Q3 of “Immune Response”) ($p$ = 0.067).
However, no such clear difference is seen for the physics questions.
In the interviews, all four students from Group L (L1-4) coincidentally explained the role of analogies based on subjects.
They noted that explanations for physical concepts are relatively concise, allowing them to understand directly without analogies. 
In contrast, the lengthy explanations for biological concepts made analogies helpful to ``\textit{get an overview and quickly identify key terms}'', as indicated by L4.

\textbf{LLM-generated analogies may negatively affect students' understanding without teacher intervention due to \chirev{errors and missing information in analogies and students' incorrect learning strategies with over-reliance}.}
\chirev{Although some students in Group L identified the analogy of ``Nuclear Fission and Fusion'' as an \textbf{Incorrect Analogy} and noted specific LLMs' hallucination during the interview, Group L's accuracy on all five questions was no higher than Group T's (Fig.~\ref{fig:study1_acc_question}).}
Furthermore, we also found that, although the \chirev{\textbf{Correct and Satisfying Analogies}} slightly improved overall answer accuracy, they could also harm students' understanding in some cases due to missing information.
For example, Group L's accuracy for Q4 of ``Immune Response'' was 0.27 lower than Group T's.
Their incorrect answer choices suggest that some students believed that ``plasma cells can recognize antigens.''
However, the textbook explains that ``antigen-presenting cells such as B cells recognize antigens'' and ``plasma cells release antibodies to eliminate antigens'', while the LLM-generated analogy only includes the latter and omits the former.
\chirev{This may be linked to over-reliance issue, as L1 and L3 described their learning strategies during the interview as ``\textit{reading the analogy first, answering the questions, and not revisiting the textbook if the answers seemed clear from the analogy.}''}


\textbf{Students subjectively appreciate the correct LLM-generated analogies often with overconfidence.}
We \chirev{were surprised to find} the strongest association between group and self-confidence was for ``Wave-Particle Duality'' ($p = 0.025$) among the four concepts. 
This suggests that students were receptive to the \chirev{\textbf{Correct Analogy with Imagination}}. 
However, there was no significant association between group and answer accuracy for this concept. 
We also observed overconfidence among \chirev{\textbf{Correct and Satisfying Analogies} for} biological concepts: a significant association between group and confidence ratings for Q2 of “Blood Sugar Regulation” ($p$ = 0.034), but none with accuracy ($p$ = 1).
As shown on the right of Fig.~\ref{fig:study1_acc_conf_overall}, there are significant associations between group and confidence ratings for both subjects.
Besides, We found a negligible correlation between accuracy and the confidence rating, as the absolute value of Kendall's tau correlation coefficient between them in each group and concept was $\leq0.2$.


\chirev{Overall, our empirical evidences suggest that LLM-generated analogies are currently unsuitable for unsupervised self-learning systems. We will discuss LLMs in supporting students' learning by analogy in Sec.~\ref{sec: discussion_student}.}

\section{Study II}
\label{sec:study2}
In this section, we introduce the pre-class interview (Sec.~\ref{sec:study21}) to gather requirements for the classroom experiments (Sec.~\ref{sec:study22}) that evaluate the actual use of LLM-generated analogies in classroom teaching.

\subsection{Pre-class Interview}
\label{sec:study21}
In this subsection, we outline the participants for our pre-class interview (Sec.~\ref{sec:study21_participants}), the procedure and stimulus (Sec.~\ref{sec:study21_procedure_and_stimuls}), and the findings and derived requirements (Sec.~\ref{sec:study21_findings_and_derived_requirments}) for further classroom experiments. 
\subsubsection{Participants}
\label{sec:study21_participants}
We recruited two \chifinal{Chinese} teachers and two \chifinal{Chinese} students from the same high school as Study I to participate in pre-class interviews. 
The two teachers (T1 and T2; 1 female) are a physics teacher with 6 years of teaching experience and a biology teacher with 3 years of teaching experience. 
Both have a bachelor's degree, are interested in AI-assisted education, and teach first-year high school courses in the semester during our study.
The two senior students (S1 and S2; 1 female) are in their third year of high school, have learned the concepts used in Study I, and have above-average grades.

\subsubsection{Procedure}
\label{sec:study21_procedure_and_stimuls}
We conducted one-on-one semi-structured online interviews with the participants via Tencent Meeting.
The student interviews lasted 40 minutes, with a \$10 gift card for each student, while the teacher interviews lasted 60 minutes, with a \$20 gift card for each teacher.

For teachers, the interview included four steps to understand the requirements of LLM-generated analogies in teaching.

\textbf{Step 1: Analogy Orientation.} 
We first presented them with common analogies in the classroom from the literature (\eg, ``light waves and water waves'', ``heart and hydraulic pump'')~\cite{oliva_teaching_2007} to orient them to analogies and ensure the terminology used during the interview.
We asked the teachers to recall the analogies they had used and encouraged them to think aloud about any experiences with analogies throughout the interview.

\textbf{Step 2: Analogy Usage Exploration.}
After that, we conducted interviews using a questionnaire primarily based on the one proposed by~\cite{oliva_teaching_2007} but modified to incorporate insights from the latest research over the past decade.
We first investigated how teachers prepare analogies, such as whether they prepare in advance or improvise and adjust in the class. 
Then we asked teachers to share the characteristics of good analogies in their opinion~\cite{gray_teaching_2021} and whether they agree with the principles we summarized from the literature in Study I. 
We also investigated whether teachers involved students in building analogies during teaching, at which step of introducing knowledge points they used analogies, and whether they followed the six-step theoretical model about analogy~\cite{richland_analogy_2015}.
Other questions covered include whether visual aids were used and whether immediate feedback was provided on students' understanding of analogies.

\textbf{Step 3: AI Usage Exploration.}
We then asked teachers about their previous experiences with AI tools like ChatGPT, their familiarity with AI-assisted teaching or self-learning. 
We also inquired about their concerns on AI performance and its application in educational settings~\cite{chen2024stugptviz,tan_more_2024}, and whether they believe AI could partially replace teachers to achieve educational goals such as mastering basic concepts, problem-solving, and developing higher-order independent thinking skills.

\textbf{Step 4: Expectations Sharing on LLM-generated Analogies.}
We then showed each teacher the analogies from Study I in their respective teaching subjects.
We asked them to evaluate each analogy's strengths, weaknesses, classroom applicability, and potential for teacher modification
Building on this, we asked teachers to share their expectations for effective AI-generated analogies.

The interviews with the students were focused on their classroom experiences rather than exploring AI usage since their role in the classroom mainly involved receiving information rather than designing analogies, and they spent most of their time without any smart devices or AI.
Initially, students were asked to recall analogies used in class. 
We then presented the ten concepts from Study I, asking students to reflect on their learning experiences. 
Next, we presented the ten analogies from Study I and asked for their feedback on their effectiveness in enhancing their understanding.

\subsubsection{Findings and Derived Requirements}
\label{sec:study21_findings_and_derived_requirments}
\begin{table*}[t]
\centering
\small
\caption{A summary of interviewing physics and biology teachers.} \label{tab:findings_in_study2_pre_class_interview}
\begin{tabular}{@{}p{0.2\linewidth}p{0.39\linewidth}p{0.38\linewidth}@{}}
\toprule
\textbf{Topic} & \textbf{Physics Teacher (T1)} & \textbf{Biology Teacher (T2)} \\ 
\midrule
\multicolumn{3}{c}{\cellcolor[gray]{0.95}{\textbf{Analogy Usage Exploration}}} \\
\midrule
Analogies Frequency & Sometimes. & Frequent. \\
\midrule
Analogies Feature & Mostly between learned concepts. & Mostly between biology and daily life. \\
\midrule
Source of Analogies & Mostly Prepared analogies between concepts. \newline A few improvised analogies with everyday lives. & Mostly prepared analogies. \newline Nearly no improvised analogies. \\
\midrule
Good Analogy Criterion & Easy to understand and free of scientific errors. & Easy to understand and related to everyday life. \\
\midrule
Agreement with Initial Principles in Study I & Partial agreement: Analogies between similar physical concepts. & Total Agreement. \\
\midrule
Analogy Explanation & Verbal explanation + imagery + teaching aids & Verbal explanation + imagery + teaching aids \\
\midrule
Analogies Usage Scenario & Often used to introduce concepts. \newline Sometimes throughout teaching. & Often used when detailing knowledge points. \\
\midrule
Agreement with the Six-step Model of Practice~\cite{richland_analogy_2015} & Acknowledges most, except for pointing out differences when introducing concepts. & Total agreement. \\
\midrule
Student Participation in Constructing Analogies & Rare. Sometimes, students offer their ideas, which might be used in the next class. & Rare. Sometimes, students prepare analogies for student-led discussions. \\
\midrule
Students Understanding Examination & Question students with ``Have you seen something similar before?'', or observe students' expressions & Students complete a few exercises during class, or question students about concept differentiation. \\
\midrule
\multicolumn{3}{c}{\cellcolor[gray]{0.95}{\textbf{AI Usage Exploration}}} \\
\midrule
Awareness and Experience with AI & Has used ChatGPT for writing papers, lesson plans, and creating images; knows about Sora. & Has used ChatGPT for tenders and personal use. \\
\midrule
Pros and Cons of AI & Pros: helps write unexpected things. \newline Cons: Needs specific questions; AI usually doesn't follow the instructions. & Pros: Provides broad ideas.  \newline No clear cons due to limited experience. \\
\midrule
Can AI Replace Teachers? & Teachers know students' learning situations, AI does not; AI-generated content needs adjustment. & AI cannot replace but complement teachers. \\
\midrule
\multicolumn{3}{c}{\cellcolor[gray]{0.95}{\textbf{Expectations Sharing on LLM-generated Analogies}}} \\
\midrule
Positive Comments on Analogies in Study I & 1. The analogies are all vivid and some of them are interesting & 1. Some analogies are similar to those used in class \newline 2. Identify analogies to try in class for concepts not usually taught with analogies. \\
\midrule
Negative Comments on Analogies in Study I & 1. Analogies don't clarify abstract concepts. \newline 2. Analogies can complicate simple concepts. \newline 3. For concepts that are tested simply, memorization is enough. \newline 4. Pictures could make some concepts clear without analogies. & 1. Analogies shouldn't reflect all but the main concepts; the rest relies on memory. \newline 2. Pictures and animations can visualize familiar organisms without analogies. \newline 3. Although rare, related concepts sometimes are used as analogies. \\
\midrule
Overall Expectations & Vivid analogies between physical concepts. & Analogies from daily life for teaching focus; \newline Interesting analogies to stimulate learning interest.\\
\bottomrule
\end{tabular}
\vspace*{-10pt}
\end{table*}
The findings of the teacher interview were summarized in Tab.~\ref{tab:findings_in_study2_pre_class_interview}.
Based on these findings and the interview with senior students, we conclude the requirements for data preparation and classroom study design as follows.

\textbf{Providing analogies to teachers during lesson preparation.}
At the beginning of the interview, both teachers clearly stated that they often use analogies in class, with most being prepared in advance. 
The physics teacher (T1) frequently referred to analogies found in teaching aids. 
The biology teacher (T2) listed key knowledge points during lesson preparation and then considered suitable analogies, drawing on personal experience and input from other veteran teachers. 
Both teachers and students claimed that students rarely participate in the construction of analogies except in student-led discussion sessions.


\textbf{Generating analogies based on subjects' characteristics and analogy needs.}
Two teachers demonstrated apparent differences in their need for and use of analogies. 
Based on their explanations, we attribute these differences to their subjects' characteristics rather than personal preferences. 
T1 frequently used analogies between concepts like ``electric field and magnetic field,'' noting the abstract nature of physics and the difficulty of finding everyday analogies. 
In contrast, T2 primarily employed interesting everyday life analogies, such as likening ``chromosome crossing over'' to ``swapping legs between classmates''. 
However, when presented with the analogy for ``blood sugar regulation'' generated in Study I, T2 suggested it could be analogized with ``thyroid hormone regulation'', as functions are related and thus easy for students to grasp. 


\textbf{Generating analogies for teaching key points and helping students focus.}
Both teachers stated that the primary goal of using analogies was to help students understand key concepts. 
Additionally, they emphasized that some analogies helped students maintain engagement. 
T1 mentioned, \textit{``When I notice students getting sleepy, I occasionally improvise an interesting analogy related to the concept to wake them up.''} 
T2 used images of people and mummies to explain the dry and fresh weight of cells, which are vivid and engaging without distracting students. 


\chirev{\textbf{Generating necessary analogies determined by teachers.}}
Both teachers acknowledged our generated vivid analogies in Study I. 
However, they criticized many of them as being overly complicated and unnecessary. 
For ``nuclear fission and fusion'' and ``auxin,'' T1 and T2 pointed out that students could quickly understand them through pictures and animations. 
Additionally, T1 mentioned that the ``molecular kinetic theory'' is relatively simple and not a key focus of exams, thus only requiring memorization.
T1 also stated that concepts in atomic physics, such as the ``photoelectric effect,'' are too isolated from other physical concepts to be conveyed through analogy. 
Additionally, students interviewed could not recall many concepts taught using analogies and viewed many analogies in Study I as redundant.

\chirev{\textbf{Generating non-complex analogies for certain aspects of the concept.}}
Both teachers emphasized the importance of analogizing only parts of a concept to keep it \chirev{correct} and easy to understand. 
\chirev{
T1 took the incorrect analogy of ``nuclear fission and fusion'' (Fig.~\ref{fig:example}) to illustrate LLMs' difficulty in generating correct physical analogies, noting that forcing analogies for all features leads to factual and semantic errors. 
He explained that physical concepts often involve multiple features, some of which, like ``chain reactions'', can be analogized (e.g., ``dominoes''), while others, such as ``mass-energy conversion'', are too abstract to find counterparts due to their basis in mathematical models.
}
\chirev{For biological analogies,} T2 recommended focusing on negative feedback in ``thyroid hormone regulation'' with an analogy like ``adjusting the temperature with an air conditioner remote control,'' while students should memorize other details. 
S2 recalled an analogy about specific details, in which the teacher compared a ``channel protein'' with a ``fire escape.''
Therefore, for complex concepts with multiple knowledge points, selecting only a specific aspect for the analogy is sufficient.


\textbf{Not necessary to generate perfect analogies.}
Teachers were lenient towards the generated analogies from Study I and managed to extract effective parts from them. 
T2 appreciated the analogy comparing ``nerve impulses'' to the ``efficient operation of stations in an express delivery system,'' though some parts were redundant. 
Additionally, T1 shared his experience using ChatGPT for lesson plans, finding it repetitive and sometimes vague but useful for providing new ideas.


\textbf{Evaluating LLM-generated analogies in class by teachers.}
Teachers had various approaches to evaluating the effectiveness of analogies in class. 
T1 asked questions like ``Have you seen something similar before?'' or observed the students' expressions, while T2 had the students answer concepts-related questions during class. 
\chirev{Additionally, two teachers expressed cautious optimism about using LLM-generated analogies with their interventions. 
T1 noted that frequently used analogies for physics were concept-based, while AI-generated ones felt more relatable to everyday life, which makes him uncertain about their actual effects. 
Besides, both T1 and T2 anticipated better classroom feedback but were unsure of the effects on students' performance on homework and exams. 
This led to a consensus on conducting a comparative experiment.}



\enlargethispage{5pt}

\subsection{Classroom Experiments}
In this subsection, we describe the participants (Sec.~\ref{sec:study22_participants}), data preparation process (Sec.~\ref{sec:study22_data_preparation}), procedure (Sec.~\ref{sec:study21_findings_and_derived_requirments}), and results analysis (Sec.~\ref{sec:study22_results_analysis}) for our classroom experiments. 
\label{sec:study22}
\subsubsection{Participants}
\label{sec:study22_participants}
In this one-week field study, participants included two teachers (T1 and T2) from the pre-class interviews and two first-year high school classes (C1 and C2) they were teaching. 
Each class had 25 students, 12 of whom were girls, and the distribution of their entrance exam scores was very similar.

\subsubsection{Data Preparation}
\label{sec:study22_data_preparation}
Teachers informed us about concepts that might require analogies in the following week of teaching.
The concepts taught by the physics teacher (T1) include average velocity and instantaneous velocity, acceleration, and infinitesimal method.
The concepts taught by the biology teacher (T2) include the various functions of proteins, the adaptation of function and structure, dehydration condensation, the formation of tertiary and quaternary structures, and protein denaturation.
Based on pre-class interviews, we identify four effective strategies to generate analogies from LLMs for classroom practice.

\begin{itemize}
    \item \textbf{Strategy 1: Analogy for Physical Concept.} For physical concepts, analogies often draw on learned physical concepts. 
    \item \textbf{Strategy 2: Analogy for Biological Concept.} For biological concepts, analogies often involve everyday objects. For example, one might use the analogy of fire escape to help understand channel protein.
    \item \textbf{Strategy 3: Vivid Analogy Generation.} Analogies should be vivid and engaging to capture students' attention. 
    \item \textbf{Strategy 4: Fine-grained Analogy Generation.} Sometimes, it is sufficient to generate analogies for just one aspect of a concept to provide a detailed explanation of that particular aspect. 
\end{itemize}









Based on the strategies outlined, we can modify the prompt in Tab.~\ref{tab:instruction_prompt} to suit the specific aspect of a concept and the requirements of teachers.
Specifically, we incorporated Strategies 1, 2, and 3 into the \textit{Principles}.
Strategy 4 was added into the \textit{Input Resource} to prevent the model from forgetting.

Following discussions with T1, we generated analogies for average and instantaneous velocity by implementing either strategy 1 or 3. 
Additionally, we generated detailed analogies for the infinitesimal method and acceleration using strategy 4. 
In biology, we produced analogies for proteins by applying either strategy 2 or strategy 3. 
Using strategy 4, we developed detailed analogies for the immune effects of proteins and the formation of tertiary and quaternary structures. 
However, two generated analogies, ``driving speed'' and ``reading speed'', were marked as non-analogies and excluded. 
Besides, the analogies generated with Strategy 1 for physical concepts were not related to other concepts, but we included them as they are vivid analogies.
We generated four physical analogies to T1 and nine biological ones to T2.

\subsubsection{Procedure}
\label{sec:study22_procedure}
In this one-week teaching, T1 and T2 used LLM-generated analogies for C1 and kept the original teaching mode for C2, with each class having 3 lessons for each subject, totaling 12 lessons.
One author attended one C1 lesson taught by T1 and one by T2, observing how teachers used analogies and student reactions without disrupting teaching.
For the remaining lessons within the week, teachers completed our provided record forms after each lesson.
The record forms asked for details on which analogies they chose while preparing for C1, any modifications made to these analogies, and reasons for not selecting others. 
Additionally, the forms inquired about how teachers assessed student feedback during or after class, any differences in feedback between C1 and C2, and whether the feedback met their expectations.
After one week, we conducted one-on-one interviews with T1 and T2, each lasting 20 minutes, to clarify any unclear details in the records, and discuss their experiences with LLM-generated analogies, students' performance, and future expectations.
Both teachers received a \$60 gift card each for their dedicated participation over the week.

\subsubsection{Results Analysis}
\label{sec:study22_results_analysis}

We report the following qualitative findings based on the record forms and interviews.

\begin{figure*}[t]
    \centering
    \includegraphics[width=1\linewidth]{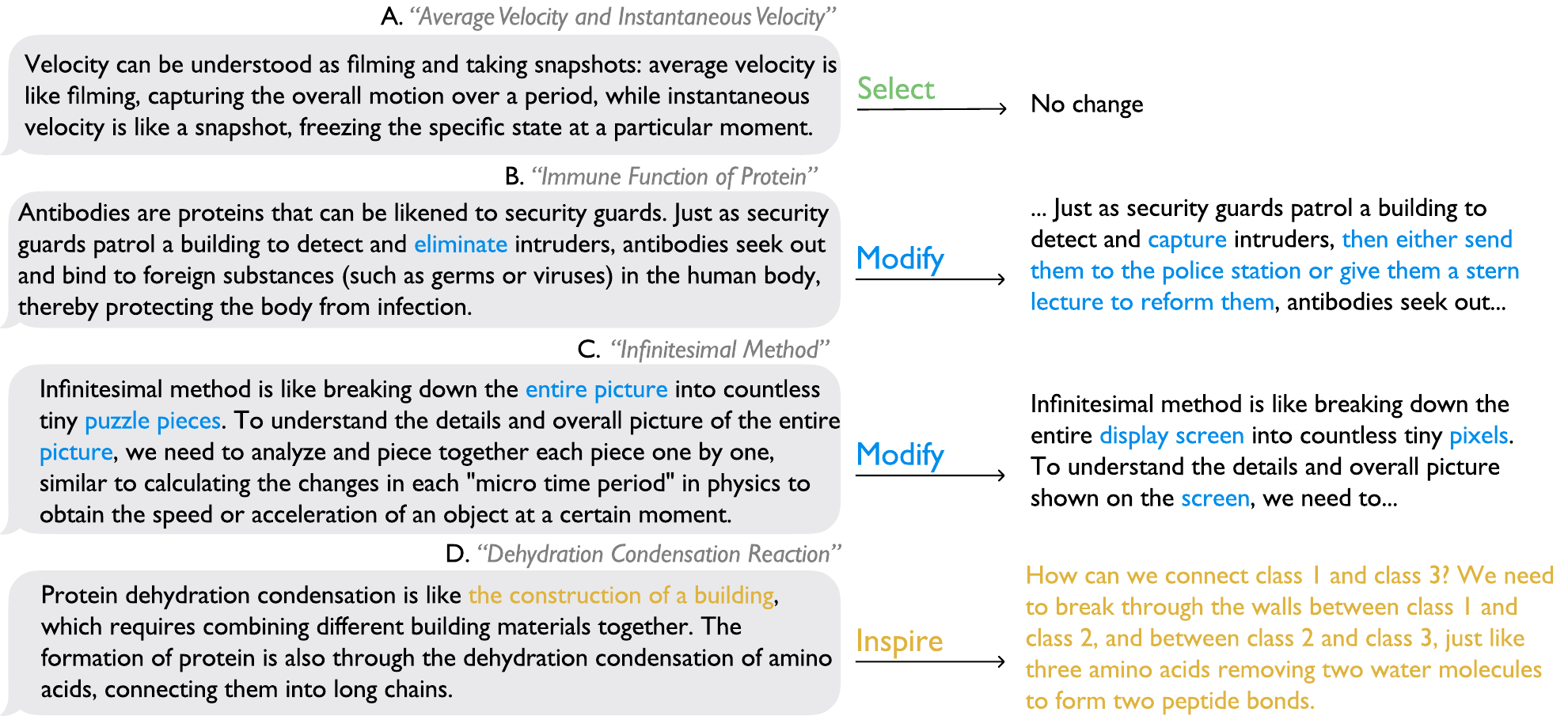}
    \caption{\chirev{Teachers' behaviors on LLM-generated analogies. They may either directly select (A) and use them in class, modify their details (B) or analogy objects (C) to varying extents, or even create entirely new analogies inspired by them (D).}}
    \Description{This figure shows examples of three types of actions that teachers can take with LLM-generated analogies. For analogies of higher quality, such as analogizing ``Average Velocity and Instantaneous Velocity'' to ``videos and snapshots'', teachers directly select and use them in class. Teachers may also modify the details of the analogy, such as changing the security guard’s ``eliminating'' the intruder in the analogy of ``Immune Function of Protein'' to ``capturing and sending him to the police station''; or modify the analogy object, such as changing the ``puzzle and puzzle pieces'' in the analogy of ``Infinitesimal Method'' to the ``entire display and a single pixel''. Teachers may even be inspired by the generated analogies to get new analogies. For example, the generated analogy of ``Dehydration Condensation Reaction'' mentioned the building structure, which inspired the teacher to get a new analogy of ``breaking down the wall between classrooms''.}
    \label{fig:behavior}
\end{figure*}

\textbf{Teachers selected and modified LLM-generated analogies to avoid redundancy, confusion, or misleading students and make them closer to students' daily lives.}
T1 selected two of four analogies and modified one, while T2 chose four of nine analogies and modified two.
In the interview, T2 noted that while the analogies for all four functions of proteins had merits, only two were selected to avoid verbosity in the class. 
Besides, to avoid concept confusion, T2 chose the analogy of ``transport function'' as a ``conveyor belt'' and discarded the analogy of ``catalysis function'' as a "high-speed elevator," due to the transport function of the elevator.
\chirev{
We observed two types of modifications made by teachers to analogies. 
One type involved modifying details, such as T2 changing the security guard's action from ``eliminating'' to ``capturing'' to align with the real-world context \chirev{(Fig.~\ref{fig:behavior}B)}.
Another type involved changing analogy objects, like T1 replacing ``jigsaw puzzle'' with ``pixels on a display screen'' to illustrate the infinitesimal method \chirev{(Fig.~\ref{fig:behavior}C)}.
T1 explained that display screens are more familiar to students than jigsaw puzzles.}

\looseness-1  \textbf{LLM-generated analogies inspire teachers with new analogies and new teaching methods.}
In the interview, both teachers recognized the novelty of some LLM-generated analogies for concepts, and they had not considered using analogies for those concepts before. 
For example, T1 used ``video and snapshot'' to analogize ``average velocity and instantaneous velocity'' \chirev{(Fig.~\ref{fig:behavior}A)}, while T2 used ``wool folding and weaving'' to analogize the ``tertiary and quaternary structures of protein.''
In addition, T2 developed new analogies inspired by LLM-generated analogies. 
While teaching the ``dehydration condensation reaction,'' T2 explained with a new analogy \chirev{as ``breaking down the wall between classrooms'' (Fig.~\ref{fig:behavior}D).}
In the interview, T2 said, ``\textit{I am not satisfied with the generated one, as comparing the dehydration condensation reaction to mixing building materials doesn't capture the essence. However, \chirev{the buldings environment} inspired me to create a new analogy.}''
Besides, T1 said that participating in this study had changed his teaching style.
T1 used analogies based on everyday life after explaining the concept, which was inconsistent with his pre-class interview response. 

\looseness-1  \textbf{LLM-generated analogies boost students' classroom and homework performance \chirev{and encourage teaching with analogy}.}
Both teachers believe that C1 outperforms C2 in both classroom participation and homework.
T1 praised analogies for helping students focus on the class: ``\textit{I can see from the students' eyes that C1 is genuinely paying closer attention, with more students nodding sincerely, rather than just pretending.}''
T1 also reported that C1 outscored C2 by nearly 20\% on a 10-question homework.
He attributed this to C2's confusion between average and instantaneous velocity, causing errors on the two hardest questions."
In the interview, T1 said, ``\textit{I plan to use the analogies in C1 when reviewing the assignments in C2 to explain the concepts again.}''
As for biological concepts, T2 said, ``\textit{When I explained protein structure using a video, C2 students understood initially but got confused about the tertiary structure, whereas the wool stacking analogy helped C1 students understand the video.}'' 
T2 showed us a fill-in-the-blank question from homework that asked students to summarize protein function. 
Most C1 students summarized correctly, while many C2 students simply copied words from the textbook. 
However, T2 noted that there was no clear difference between the two classes in understanding straightforward concepts like ``Protein denaturation''.
T1 also noted that the physics analogies are still not between concepts and may offer limited help with highly abstract concepts\chirev{, while he added ``\textit{But I'll try more teaching with analogy since the difference between the two classes is clear.''}}

\chirev{Overall, promising feedback from teachers and classroom practice led us to consider designing a practical system to support the preparation of teaching analogies.}


\enlargethispage{5pt}

\chirev{
\section{System}
\label{sec:system}
In this section, we transformed key study findings into an LLM-assisted system for teachers and conducted a system evaluation, highlighting its contribution to teaching by analogy in education.

\subsection{System Design}
This subsection details the interview for deriving design requirements and system workflow as below.

\subsubsection{Interview}
Given that Study II showed teachers prepared analogies during lesson planning, we first determined the system's necessity and functionality through 20-minute one-on-one online interviews with T1 and T2 via Tencent Meeting. 
The interview mainly consists of two questions. 1) Necessity: Is providing an LLM-assisted analogy generation system necessary for teachers to prepare lessons? 2) Functionality: What functions do they expect?

Both teachers affirmed the first question, expressing a desire to operate the system themselves to gain hands-on experience and long-term support from LLMs.
Regarding the second question, both teachers expressed a preference for the analogy generation mode in Study II. 
They suggested that after specifying the required concepts, the system should generate accurate analogies tailored to their needs, allowing for refinement and management. 
They also noted that in Study II, analogies function as plug-and-play modules to replace or enhance original explanations of scientific concepts, so the generation process need not account for other lesson plan content at this stage.

\subsubsection{Design Requirements}
After confirming the necessity and functionality, we identified three design requirements as below and confirmed them with T1 and T2.

\textbf{R1: It should incorporate principles and strategies identified in previous studies to help generate accurate analogies tailored to teacher needs.}
The general principles identified in Study I (Tab.~\ref{tab:instruction_prompt}) should be integrated into the prompt by default to enhance the accuracy.
The system should allow teachers to select useful prompting strategies identified in Study II and incorporate them into generation following the practice of data preparation of Study II (Sec.~\ref{sec:study22_data_preparation}) to better tailor analogies to their needs.

\textbf{R2: It should enable teachers to input their expectations or automatically generate personalized principles for creating analogies.}
As identified in Study II interviews (Sec.~\ref{sec:study21_findings_and_derived_requirments}), the workflow should allow teachers to input concepts they believe require analogies to be generated.
Besides, to ensure personalized needs, it should also accept user-inputted new principles tailored to their needs and even automatically generate tailored principles from user comments on generated analogies for the prompt (Tab.~\ref{tab:instruction_prompt}).

\textbf{R3: It should allow users to make manual changes and feedback to manage their analogies.}
In Study II, teachers showed a strong willingness and ability to refine and manage generated analogies (Sec.~\ref{sec:study22_results_analysis}), highlighting the need for a system that allows direct editing.
As prompt evolution in Study I showed limited improvement (Tab.~\ref{tab:error_rates}), and teachers found conversational refinement challenging in pre-class interviews in Study II (Tab.~\ref{tab:findings_in_study2_pre_class_interview}), conversational modifications by LLMs are unnecessary.
The system should also enable teachers to manage analogies by classifying them into four categories (useful, inspiring, refinable, and useless) and storing all except the useless ones.



\subsubsection{System Workflow}
\label{sec:workflow}

\begin{figure*}[t]
    \centering
    \includegraphics[width=1\textwidth]{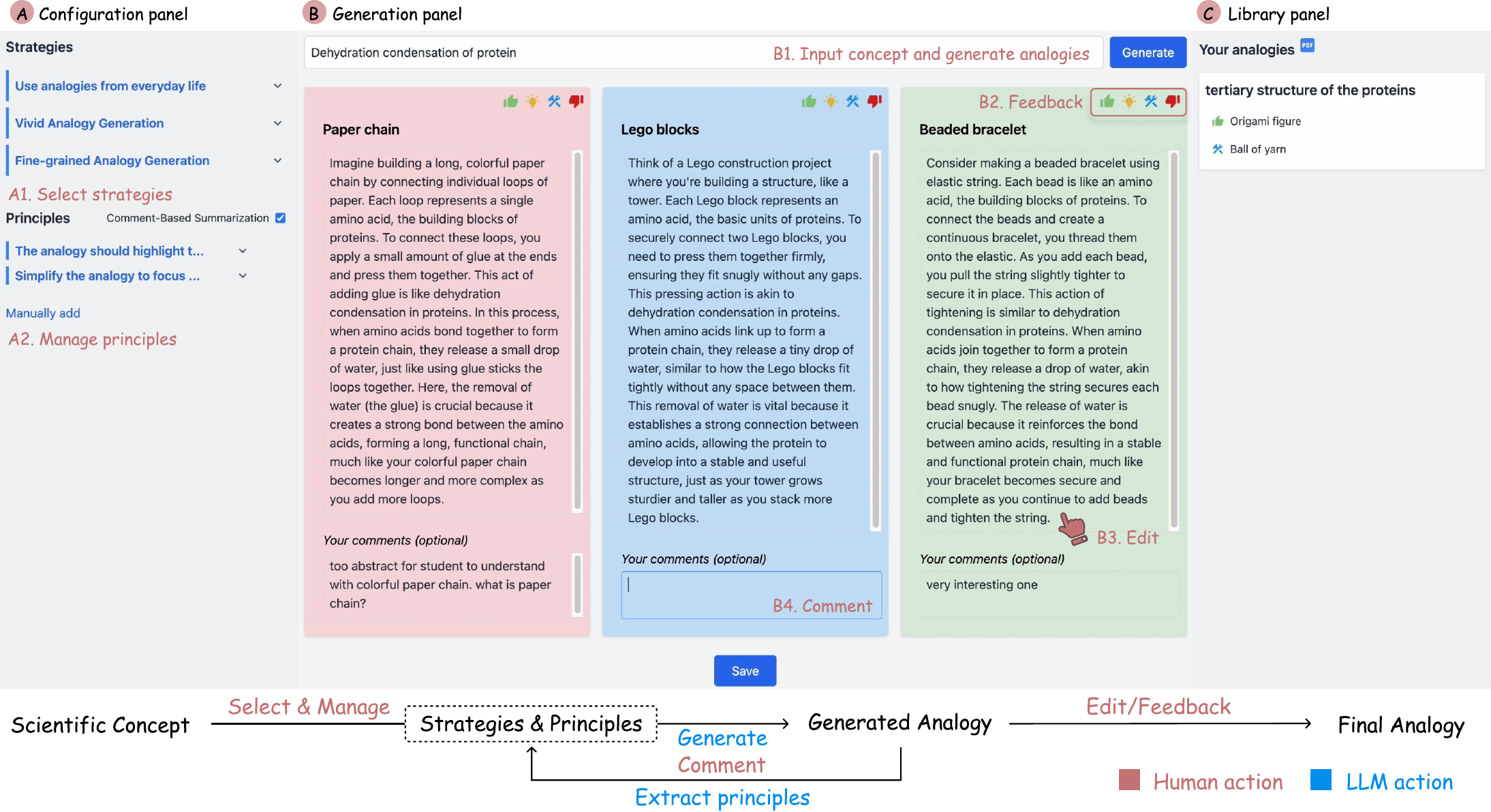}
    \caption{\chirev{Our system interface (top) and workflow (bottom). In each round, teachers use Configuration Panel (A) to select strategies (A1) and manage principles (A2) for generation. After entering scientific concepts (B1), they provide feedback (B2), edit (B3), and comment (B4) to each generated analogy in Generation Panel (B). Clicking the ``Save'' button makes the system generate new principles by LLM in Configuration Panel and store approved analogies in Library Panel (C), where teachers may export saved analogies.}}
    \label{fig:system}
    \Description{This figure shows our system interface which allows teachers to interact with three main panels. In Configuration Panel, teachers select strategies and manage principles for analogy generation. They then enter scientific concepts, provide feedback, make manual editing, and add comments to the generated analogies in Generation Panel. Clicking the ``Save'' button enables the system to generate new principles in Configuration Panel and store approved analogies in Library Panel, where all saved analogies can be exported at any time.
    }
\vspace*{-10pt}
\end{figure*}

Based on the design requirements, we built an LLM-assisted system (Fig.~\ref{fig:system}) for teachers to create and refine analogies for teaching. 
Teachers begin by registering an account and follow a workflow as below.

\textbf{Configure strategies and principles for analogy generation.} 
After selecting a teaching subject during registration, the system provides prompting strategies in Configuration Panel (Fig.~\ref{fig:system}A) based on the chosen subject. 
Teachers can click on these strategies for generating analogies (\textbf{R1}).
The principle list starts blank, allowing teachers to manually add or select principles as needed (\textbf{R2}). 
When the user hovers over a principle, they could edit or delete it freely.

\textbf{Generate analogies and provide feedback}. Teachers then input a scientific concept in Generation Panel (Fig.~\ref{fig:system}B) and click the ``Generate'' button (\textbf{R2}). 
The system will provide three cards with generated analogies. 
Each card has four feedback options in the top-right corner: useful, inspiring, refinable, and useless. 
Teachers need to select one feedback option for each analogy card and may edit analogies as needed and provide text comments (\textbf{R3}).

\textbf{Save analogies, optionally generate new principles, and restart.} When saving, modified analogies (excluding those marked as useless) are stored in Library Panel (Fig.~\ref{fig:system}C), along with the corresponding concept (\textbf{R3}). 
If teachers enable the ``Comment-based summarization'' feature in Configuration Panel (Fig.~\ref{fig:system}A), the system will summarize new principles from comments on the three analogies by LLMs and add them to the principles list (\textbf{R2}). 
Teachers can edit or delete any generated principles or disable the automatic summarization feature at any time.
After saving, the system clears the concepts and analogies in Generation Panel (Fig.~\ref{fig:system}B), allowing users to restart. 
Teachers can reconfigure the strategy list, and add, delete, select, or modify the principle list for the new round.

Besides, teachers click right buttons of text for strategies (Fig.~\ref{fig:system}A1), principles (Fig.~\ref{fig:system}A2), and approved analogies (Fig.~\ref{fig:system}C) to view or collapse text, reducing clutter. 
All approved analogies stored in Library Panel can be exported to PDF for teachers' usage.

We implemented the system as a Vue-based web application with a Python Flask backend. More implementation details, such as prompts for automatically generated principles, are provided in the supplementary material.

\subsection{System Evaluation}
This subsection details the participants, procedure, and findings of the system evaluation.
\subsubsection{Participants}
We invited 6 high school physics and biology teachers from 5 schools, including T1 and T2, along with 4 new teachers from different schools to increase diversity. 
The physics teachers had 7 years (T1), 3 years (T3), and 4 years (T5) of experience, while the biology teachers had 4 years (T2), 2 years (T4), and 25 years (T6) of experience. All participants received a \$20 gift card as compensation.

\subsubsection{Procedure}
Our study consists of a tutorial, free exploration, one-week usage, and an interview.

\textbf{Tutorial.} We conducted one-on-one online interviews with each teacher other than T1 and T2 via Tencent Meeting, lasting 20 minutes. 
We briefly revisited Steps 1 and 3 from Study II (Sec.~\ref{sec:study21}) to understand their experience with analogical teaching and AI, and to ensure they were familiar with the study background.

\textbf{Free exploration.} Following the tutorial, we continued the interview with each teacher to demonstrated the basic functionality of the system according to the workflow, after which the teachers were encouraged to explore the system freely. 
During this exploration phase, we prompted them to think aloud and we answered any questions they had to make sure they understand how to use the system. 
This process lasted about 20 minutes. 

\textbf{One-week usage.} Before using the system, we informed the teachers that it would collect data on the analogies and principles they generated, as well as their feedback, manual edits, comments, and other interaction data for research purposes. 
All of them provided informed consent. 
We then asked the teachers to use the system to generate analogies to support their lesson preparation for one week.
The frequency of system use and the content of lesson preparation were determined by the teachers themselves, although we encouraged them to use the system frequently.

\textbf{Interview.}
After one-week usage, we conducted one-on-one semi-structured online interviews with each teacher via Tencent Meeting, lasting nearly 40 minutes each. 
During the interviews, we asked each teacher the following questions: course coverage (Q1), system usability (Q2), satisfaction with the generated analogies (Q3), satisfaction with the generated principles (Q4), satisfaction with the edited analogies (Q5), satisfaction with the edited principles (Q6), and the significance of the system for their future lesson preparation and teaching (Q7).




\enlargethispage{10pt}

\subsubsection{Findings}
\label{sec:system_findings}
Based on user interaction data and interview results of Q1-Q7, we summarize the following findings.

\textbf{The system's usability and the usefulness of generated analogies were well-received by teachers (Q1-Q3).} 
Teachers reported that they used the system to generate analogies for lessons spanning from one month to half a semester, aiding both reflection on past lessons and preparation for upcoming ones. 
All teachers agreed that the system was easy to use, even if they had nearly no experience using AI (T3, T5, T6).
They generated 15 to 42 analogies (15, 15, 21, 24, 24, 42), with 40\% to 80\% marked as non-useless (``useful'', ``refinable'', ``inspiring'') across users (40\%, 58.3\%, 61.9\%, 64.3\%, 66.7\%, 80\%). 
They all agreed that although many of the generated analogies had various issues, overall, they helped expand their lesson planning ideas and inspired greater use of analogies in teaching.
For example, the system analogizes ``phase difference'' to ``some students doing radio gymnastics faster or slower than classmates.'' T3 said, ``\textit{I hadn’t thought of that, but it’s great, and since my class is right after gymnastics session, I’ll definitely use it.}''
Besides, physics teachers noted that they understood the system’s inability to provide suitable analogies for some complex concepts, as such analogies might not exist anyway.

\textbf{Teachers tried to improve analogy quality by incorporating generated principles or directly regenerating analogies (Q4).} 
Five teachers (excluding T4) continuously enabled the automatic principle generation feature and actively input their comments, producing 10 to 30 principles (10, 13, 17, 18, 30) and incorporating at least half into analogy generation.
They all praised the quality of the principles, with T1 saying, ``\textit{The system infers general principles from my vague intents on comments on specific analogies.}''
For usability of this feature, T3 noted, ``\textit{It doesn’t matter if too many principles are generated during the usage. I just delete the redundant ones—it’s more convenient than summarizing myself.}'' 
For its effectiveness in analogy generation, T1 and T5 felt the principles improved analogy quality, while T3 and T6 were unsure but still incorporated them because ``\textit{adding more correct terms can't hurt.}''
On the contrary, T4 preferred regenerating analogies directly without commenting on analogies and generating principles, saying, ``\textit{If the analogies aren't good, I just regenerate them a few more times as I know the randomness of AI.}''
Only T4 manually added principles at the beginning of usage, while all teachers found it inconvenient and difficult as noted by T3 and T1.

\textbf{The generated analogies and principles benefit teachers more than just concept explanations (Q7).} 
1) The principles help shape teaching expertise. 
T1 said, ``\textit{I can learn teaching techniques from the generated principles, and after using the system for a while, I could even write a teaching paper. It’s like having a discussion about teaching with another experienced teacher. While the principles may not immediately impact teaching, the long-term accumulation is valuable.}'' 
2) The analogies supplement the teacher's knowledge base. 
T6 noted that the generated analogies broaden a teacher's perspective, and saving more analogies gradually builds their teaching knowledge system, which help teachers adapt to different teaching situations. 
3) The analogies also inspire quiz and test creation.
T2 said, ``\textit{Even if some analogies aren't ideal, I save them because having students identify errors helps their learning.}'' T3 stated, ``\textit{Some of the generated `analogies' is example-based explanation, but I save it because these real-life examples can be used to set questions.}''

\textbf{Teachers typically organize analogies externally rather than refining them within the system and there is potential over-reliance (Q5-Q6).}
Despite marking large proportion of generated analogies as ``refinable'' (Mean = 21.4\%) and ``inspiring'' (Mean = 13.8\%), only T1 and T4 edited 1 analogy in the system, respectively. 
T6 explained, ``\textit{Manually modifying so much text is too burdensome for older teachers.}'' 
However, all teachers reported improving analogies to fit their needs through external modifications.
T5 noted that his lesson preparation habit is to record keywords in electronic notes, so the analogies in the system only serve as explanation and inspiration, which he then reorganizes in his notes.
Similarly, T1 and T6 preferred recording analogies in paper notebooks. 
This separation between analogy generation and actual lesson preparation may make it difficult to supervise teachers’ behaviors in teaching and potentially lead to teachers' over-reliance on generated analogies.
This issue may be more pronounced for users like T4, who prefer to regenerate analogies directly without providing any comment, compared to teachers who actively engage by entering comments in the system.

}

\section{Discussion}
\label{sec:discussion}
\chirev{This section discusses consideration, opportunities, and future research directions for LLM-assisted analogical education based on our study results and designed system.}

\chifinal{
\subsection{Subject Differences in LLM-Generated Analogy Effectiveness}

Our study shows that LLMs generally produce correct and satisfying analogies for biological concepts but generate incorrect or correct yet unprofessional ones for physics.

Several factors appear to contribute to these shortcomings. 
First, physical concepts are highly abstract (e.g., ``mass-energy conversion''), with complex and formula-driven features, making it difficult to find real-life analogies or other concepts that perfectly align with them. 
In contrast, biological concepts are more concrete and observable, often tied to specific structures and functions (e.g., ``mitochondria as the powerhouse of the cell'').
As a result, LLMs would produce forced analogies or oversimplifications for physics while generating satisfying ones for biology.
Second, restricting the physics analogy to a single aspect, as in the strategies used in Study II (Sec.~\ref{sec:study22_data_preparation}), can yield correct and engaging analogies.
However, for highly abstract concepts, these analogies may still be superficial and offer limited support, as noted by physics teachers in Sec.~\ref{sec:study22_results_analysis}.
Third, teaching materials in physics contain more formulas and fewer analogies compared to biology. As a result, LLMs learn fewer physics analogies and generate less effective analogies.

These findings suggest that using LLMs for educational analogy generation is tied to subject characteristics, and we can infer that it may be particularly challenging for subjects lacking clear real-world counterparts (e.g., mathematics). 
In contrast, they might work better for subjects with more directly observable phenomena  (e.g., high school chemistry, biology), which should be confirmed by future studies. 
Nonetheless, analogies help students engage with abstract subjects like physics and math by inspiring interest and sustaining attention. 
More studies are needed to verify the effectiveness and needs of LLM-generated analogies across a broader range of subjects, in conjunction with the review and refinement of teachers before their use.
}

\chirev{
\subsection{Generating High-Quality Analogies}
The automatically generated analogies have limitations in scientific accuracy and educational effectiveness. 
In our practical system, LLMs summarized principles from human feedback and incorporated them into the next round of analogy generation.
Several teachers in the system evaluation found this approach effective for improving analogy quality.
Building on this, we can incorporate Reinforcement Learning with Human Feedback (RLHF). 
Using teachers' feedback and preferences, we can create a reward model that continually refines the analogies.
To further reduce teachers' workload, future work should explore automatic methods for generating higher-quality analogies.
\chifinal{
First, we could explore multi-agent collaboration methods to further mitigate hallucinations~\cite{SHI2025125723}, including factuality errors and consistency errors, as outlined in Study I.
Besides}, instead of waiting for more advanced general-purpose LLMs to be released~\cite{openai_o1_preview_2024}, we can fine-tune existing models with teacher-adjusted analogies.
Additionally, our system can be transformed into a labeling tool to collect high-quality educational analogy datasets, consisting of revised analogies or new analogies proposed by teachers.
Another bottleneck for generating high-quality analogies is the LLMs' limited domain-specific knowledge. 
In Study II, the physics teacher attributed the current analogies' interesting yet unprofessional nature to the LLMs' limited understanding of abstract physics concepts.
To address this, we should fine-tune LLMs for specific subjects, improving their understanding and enhancing analogy quality.
Finally, \chifinal{to improve control over the complexity and ethical considerations and make analogies suited for the intended educational level and scenario,} the future analogy generation pipeline should consider factors like students' educational background, cultural context, and prior knowledge.
}

\chirev{
\subsection{LLMs \chifinal{for} Teaching by Analogy}
Our Study II and system evaluation show that LLM-generated analogies are valuable to teachers in three progressive aspects: short-term lesson preparation, teaching strategy development, and professional growth. 
For short-term lesson preparation, teachers are able and willing to select and modify LLM-generated analogies or inspire new ones to suit specific concepts and lessons (Sec.~\ref{sec:system_findings}), or use generated content to help set quizzes (Sec.~\ref{sec:study22_results_analysis}). 
For teaching strategy development, continuous use of LLM-generated analogies leads to positive feedback from the classroom and students' homework and iteratively encourages teaching by analogy (Sec.~\ref{sec:study22_results_analysis}). 
Regarding professional growth (Sec.~\ref{sec:system_findings}), providing feedback on LLM-generated analogies helps teachers actively reflect on teaching points and build their knowledge base, while LLM-generated principles based on their feedback also serve as valuable reminders for teachers, supporting their ongoing professional development and enhancing teaching expertise.
Given these benefits, future work should explore the varying needs and develop practical systems to benefit teachers with different experience levels and subjects.
In addition, long-term evaluation of such practical systems and teachers is needed to fully understand the actual benefits.

Besides the benefits, over-reliance on LLM-generated analogies warrants attention. 
In Study II, teachers emphasized during pre-class interviews and demonstrated in classroom teaching that they could avoid over-relying on such content.
However, in the system evaluation, most teachers did not revise analogies within the system but recorded changes elsewhere, following their own lesson preparation habits. 
This suggests that monitoring teachers' interactions with analogies only through system logs might be insufficient, potentially allowing unnoticed over-reliance to develop.
To address this, the system could use pop-up reminders to alert users against over-reliance and encourage them to edit or provide comments when no manual interaction is detected for an extended period.
Nevertheless, supervision from schools and higher authorities is essential. Additionally, regular updates from system developers to educators~\cite{tan_more_2024,kasneci2023chatgpt} are crucial for maintaining an accurate understanding of model capabilities and ensuring effective use.
}

\chirev{
\subsection{\chifinal{Integrating Analogies into LLM-Assisted Education Platforms}}
\label{sec: discussion_student}

\chifinal{
Integrating analogy generation into LLM-assisted education platforms might benefit teachers and students.

For teachers, the interplay between analogy generation and LLM-assisted teaching material preparation is mutually reinforcing.
First, analogies help develop teaching materials by providing relatable explanations.
For example, LLM-assisted platforms already help novice teachers generate lesson plans\chirev{~\cite{lessonplanner2024uist}}. 
Integrating analogy generation into these platforms can support analogy-based explanations at different teaching stages.
Additionally, participants in our system evaluation demonstrated the potential of using analogies in quiz settings, highlighting its role in automated problem generation.
Second, existing LLM-assisted teaching preparation platforms enhance context-aware analogy generation and verification. 
These platforms usually consider students’ knowledge levels and course context~\cite{GeneratingContextualized}, which could be integrated into analogy generation pipelines to help generate analogies suited for practical scenarios. 
Moreover, existing problem generation platforms could be enhanced to generate quizzes that verify students' understanding of analogy, reinforcing the effectiveness of teaching by analogy.
}

\chifinal{However, the} results of Study I indicate that LLM-generated analogies without human intervention are unreliable for students, making it premature to directly integrate analogy generation into self-learning systems. 
In contrast, teacher-adjusted analogies in Study II ensured correctness and reliably impacted students' classroom feedback and homework performance. 
Therefore, future self-learning systems should only consider pre-set teacher-reviewed analogies for key knowledge points to aid student understanding. 
However, even correct biological analogies from Study I led to negative effects, with students over-relying on them with incorrect learning strategies or becoming subjectively overconfident.
This suggests that the self-learning system should flexibly structure the learning process and, after presenting analogies, guide students back to the textbook with more detailed follow-up questions.
Besides, timely pre-set and teacher-reviewed exercises with feedback and explanation that reveal the limitations of analogies help students reflect on their learning and monitor their learning approaches.
Overall, incorporating analogies into self-learning systems requires careful attention from teachers and system developers to mitigate potential negative effects.
}

\subsection{Evaluating Broader Analogies in Education}
Analogies serve multiple purposes beyond students' understanding \chirev{and teachers' teaching}, which adds challenges to their evaluation.
In mathematical problems and similar domains, specific procedures involving numeracy and variables often require a different type of analogy, known as procedural analogy~\cite{richland_analogy_2004}. 
Such procedural analogies were also mentioned during our interview with the physics teacher in Study II. 
Due to their rarity and complexity, these analogies, even those crafted by humans, have not been thoroughly evaluated.
LLMs can lower the barrier to creating such analogies due to advanced reasoning abilities and broad subject knowledge, such as linking the formula for a ``spring oscillator'' with that of a ``pendulum.'' 
Our study design can be extended to such analogies by involving calculation questions involving formulas in controlled in-class tests.
Additionally, analogies are utilized for socialization, helping to educate children on becoming better students and enacting behavioral changes~\cite{richland_analogy_2004}.
The evaluation of such analogies involves contexts beyond the classroom, which brings challenges to study design and needs to be explored in the future.

\subsection{Limitation}
Although we have gained lots of evidence and knowledge, our work is limited by student and teacher participation and analogy representation.



\subsubsection{Limitations in Student and Teacher Participation.}
\chirev{Due to practical limitations, the teachers and students in our first two studies were from one high school, and the sample size was limited.
To explore real-world practicality, we expanded the participant pool by inviting more teachers from different schools with varying teaching experiences to evaluate the practical system.
Additionally, our study is limited to physics and biology due to practical constraints, excluding other subjects like chemistry.
Besides, our study with high school students may not be generalizable to younger students who might not possess developed analogical reasoning abilities or have limited world knowledge~\cite{vendetti_analogical_2015}.
\chifinal{Moreover, our study's demographic generalizability is limited, as all participants are Chinese, while prior research~\cite{richland_cognitive_2007} suggests that U.S. teachers provide cognitive support for analogies less frequently than teachers from Hong Kong or Japan in math instruction.}
In future work, we intend to expand the sample size, range, and diversity of subjects and explore diverse education levels for comprehensive evaluation.}

\subsubsection{Limitations in Representation of Analogy.}
The practical use of analogies in teaching extends beyond the free-form analogies we generated, incorporating visual aids and dynamic technologies to enhance understanding and interaction~\cite{richland_cognitive_2007,richland_analogy_2015}.
Our interviews in Study II \chirev{and system evaluation} also revealed that teachers \chirev{have the desire to} use images, videos, and physical aids to convey analogies. 
Moving forward, we plan to enrich LLM-generated analogies with rich text, structured representations, and generated visuals to benefit teachers \chirev{in practical systems.}



\section{Conclusion}
\label{sec:conclusion}
\chirev{
In this work, we first conducted in-class tests and classroom experiments guided by pre-class interviews to evaluate the effectiveness of LLM-generated analogies in two educational scenarios. 
Our in-class tests suggest that LLM-generated analogies could be beneficial for students' understanding, especially on biology concepts, but are unsuitable for self-learning systems without teacher intervention due to students' over-reliance and overconfidence. 
Classroom experiments reveal that teachers effectively create or refine analogies to meet their needs and are encouraged to teach with analogy by positive student feedback in class and homework.
Building on these findings, we developed a practical system for teachers preparing analogies.
Teachers in the system evaluation recognize its real-world effectiveness in lesson preparation about concept explanation and quiz design, teaching methods, and professional growth, despite potential over-reliance issues.
We hope future tool designers could consider these factors to ensure that LLM-generated analogies have a successful impact on teaching and learning.
}

\begin{acks}
The authors want to thank the reviewers for their suggestions. 
The authors also thank SHANGHAI CUNZHI HIGH SCHOOL for its generous support and all participating teachers and students, especially Siwei Ye and Zihan Luo for their deep involvement, and Yuhui Wang for strong support.
This work is supported by Natural Science Foundation of China (NSFC No.62472099, No.62202105, and No.92270121).
\end{acks}

\balance
\bibliographystyle{ACM-Reference-Format}
\bibliography{sample-base}


\begin{thebibliography}{88}


\ifx \showCODEN    \undefined \def \showCODEN     #1{\unskip}     \fi
\ifx \showDOI      \undefined \def \showDOI       #1{#1}\fi
\ifx \showISBNx    \undefined \def \showISBNx     #1{\unskip}     \fi
\ifx \showISBNxiii \undefined \def \showISBNxiii  #1{\unskip}     \fi
\ifx \showISSN     \undefined \def \showISSN      #1{\unskip}     \fi
\ifx \showLCCN     \undefined \def \showLCCN      #1{\unskip}     \fi
\ifx \shownote     \undefined \def \shownote      #1{#1}          \fi
\ifx \showarticletitle \undefined \def \showarticletitle #1{#1}   \fi
\ifx \showURL      \undefined \def \showURL       {\relax}        \fi
\providecommand\bibfield[2]{#2}
\providecommand\bibinfo[2]{#2}
\providecommand\natexlab[1]{#1}
\providecommand\showeprint[2][]{arXiv:#2}

\bibitem[Alkhatib and Bernstein(2019)]%
        {streetlevel2019Alkhatib}
\bibfield{author}{\bibinfo{person}{Ali Alkhatib} {and} \bibinfo{person}{Michael
  Bernstein}.} \bibinfo{year}{2019}\natexlab{}.
\newblock \showarticletitle{Street-Level Algorithms: A Theory at the Gaps
  Between Policy and Decisions}. In \bibinfo{booktitle}{\emph{Proceedings of
  the 2019 CHI Conference on Human Factors in Computing Systems}} (Glasgow,
  Scotland Uk) \emph{(\bibinfo{series}{CHI '19})}.
  \bibinfo{publisher}{Association for Computing Machinery},
  \bibinfo{address}{New York, NY, USA}, \bibinfo{pages}{1–13}.
\newblock
\showISBNx{9781450359702}
\urldef\tempurl%
\url{https://doi.org/10.1145/3290605.3300760}
\showDOI{\tempurl}


\bibitem[Bhavya et~al\mbox{.}(2022)]%
        {bhavya_analogy_2022}
\bibfield{author}{\bibinfo{person}{Bhavya Bhavya}, \bibinfo{person}{Jinjun
  Xiong}, {and} \bibinfo{person}{ChengXiang Zhai}.}
  \bibinfo{year}{2022}\natexlab{}.
\newblock \showarticletitle{Analogy {Generation} by {Prompting} {Large}
  {Language} {Models}: {A} {Case} {Study} of {InstructGPT}}. In
  \bibinfo{booktitle}{\emph{Proceedings of the 15th {International}
  {Conference} on {Natural} {Language} {Generation}}}.
  \bibinfo{publisher}{Association for Computational Linguistics},
  \bibinfo{address}{Waterville, Maine, USA and virtual meeting},
  \bibinfo{pages}{298--312}.
\newblock
\urldef\tempurl%
\url{https://aclanthology.org/2022.inlg-main.25}
\showURL{%
\tempurl}


\bibitem[Bhavya et~al\mbox{.}(2024)]%
        {bhavya2024analego}
\bibfield{author}{\bibinfo{person}{Bhavya Bhavya}, \bibinfo{person}{Yang Zhou},
  \bibinfo{person}{Shradha Sehgal}, \bibinfo{person}{Suma Bhat}, {and}
  \bibinfo{person}{ChengXiang Zhai}.} \bibinfo{year}{2024}\natexlab{}.
\newblock \showarticletitle{Analego: Let's build analogies together!}. In
  \bibinfo{booktitle}{\emph{AI for Education: Bridging Innovation and
  Responsibility at the 38th AAAI Annual Conference on AI}}.
\newblock


\bibitem[Boteanu and Chernova(2015)]%
        {boteanu_solving_2015}
\bibfield{author}{\bibinfo{person}{Adrian Boteanu} {and} \bibinfo{person}{Sonia
  Chernova}.} \bibinfo{year}{2015}\natexlab{}.
\newblock \showarticletitle{Solving and {Explaining} {Analogy} {Questions}
  {Using} {Semantic} {Networks}}.
\newblock \bibinfo{journal}{\emph{Proceedings of the AAAI Conference on
  Artificial Intelligence}} \bibinfo{volume}{29}, \bibinfo{number}{1}
  (\bibinfo{date}{Feb.} \bibinfo{year}{2015}).
\newblock
\urldef\tempurl%
\url{https://doi.org/10.1609/aaai.v29i1.9400}
\showDOI{\tempurl}


\bibitem[Brown et~al\mbox{.}(1989)]%
        {brown_analogical_1989}
\bibfield{author}{\bibinfo{person}{Ann~L. Brown}, \bibinfo{person}{Mary~Jo
  Kane}, {and} \bibinfo{person}{Carolyn Long}.}
  \bibinfo{year}{1989}\natexlab{}.
\newblock \showarticletitle{Analogical transfer in young children: {Analogies}
  as tools for communication and exposition}.
\newblock \bibinfo{journal}{\emph{Applied Cognitive Psychology}}
  \bibinfo{volume}{3}, \bibinfo{number}{4} (\bibinfo{year}{1989}),
  \bibinfo{pages}{275--293}.
\newblock
\showISSN{1099-0720}
\urldef\tempurl%
\url{https://doi.org/10.1002/acp.2350030402}
\showDOI{\tempurl}
\newblock
\shownote{\_eprint:
  https://onlinelibrary.wiley.com/doi/pdf/10.1002/acp.2350030402}.


\bibitem[Casarett et~al\mbox{.}(2010)]%
        {casarett_can_2010}
\bibfield{author}{\bibinfo{person}{David Casarett}, \bibinfo{person}{Amy
  Pickard}, \bibinfo{person}{Jessica~M. Fishman}, \bibinfo{person}{Stewart~C.
  Alexander}, \bibinfo{person}{Robert~M. Arnold}, \bibinfo{person}{Kathryn~I.
  Pollak}, {and} \bibinfo{person}{James~A. Tulsky}.}
  \bibinfo{year}{2010}\natexlab{}.
\newblock \showarticletitle{Can {Metaphors} and {Analogies} {Improve}
  {Communication} with {Seriously} {Ill} {Patients}?}
\newblock \bibinfo{journal}{\emph{Journal of Palliative Medicine}}
  \bibinfo{volume}{13}, \bibinfo{number}{3} (\bibinfo{date}{March}
  \bibinfo{year}{2010}), \bibinfo{pages}{255--260}.
\newblock
\showISSN{1096-6218}
\urldef\tempurl%
\url{https://doi.org/10.1089/jpm.2009.0221}
\showDOI{\tempurl}


\bibitem[Chan et~al\mbox{.}(2018)]%
        {solvent2018chan}
\bibfield{author}{\bibinfo{person}{Joel Chan}, \bibinfo{person}{Joseph~Chee
  Chang}, \bibinfo{person}{Tom Hope}, \bibinfo{person}{Dafna Shahaf}, {and}
  \bibinfo{person}{Aniket Kittur}.} \bibinfo{year}{2018}\natexlab{}.
\newblock \showarticletitle{SOLVENT: A Mixed Initiative System for Finding
  Analogies between Research Papers}.
\newblock \bibinfo{journal}{\emph{Proc. ACM Hum.-Comput. Interact.}}
  \bibinfo{volume}{2}, \bibinfo{number}{CSCW}, Article \bibinfo{articleno}{31}
  (\bibinfo{date}{nov} \bibinfo{year}{2018}), \bibinfo{numpages}{21}~pages.
\newblock
\urldef\tempurl%
\url{https://doi.org/10.1145/3274300}
\showDOI{\tempurl}


\bibitem[Chen et~al\mbox{.}(2022)]%
        {chen_e-kar_2022}
\bibfield{author}{\bibinfo{person}{Jiangjie Chen}, \bibinfo{person}{Rui Xu},
  \bibinfo{person}{Ziquan Fu}, \bibinfo{person}{Wei Shi},
  \bibinfo{person}{Zhongqiao Li}, \bibinfo{person}{Xinbo Zhang},
  \bibinfo{person}{Changzhi Sun}, \bibinfo{person}{Lei Li},
  \bibinfo{person}{Yanghua Xiao}, {and} \bibinfo{person}{Hao Zhou}.}
  \bibinfo{year}{2022}\natexlab{}.
\newblock \showarticletitle{E-{KAR}: {A} {Benchmark} for {Rationalizing}
  {Natural} {Language} {Analogical} {Reasoning}}. In
  \bibinfo{booktitle}{\emph{Findings of the {Association} for {Computational}
  {Linguistics}: {ACL} 2022}}. \bibinfo{publisher}{Association for
  Computational Linguistics}, \bibinfo{address}{Dublin, Ireland},
  \bibinfo{pages}{3941--3955}.
\newblock
\urldef\tempurl%
\url{https://doi.org/10.18653/v1/2022.findings-acl.311}
\showDOI{\tempurl}


\bibitem[Chen et~al\mbox{.}(2024a)]%
        {bidtrainer2024chen}
\bibfield{author}{\bibinfo{person}{Liuqing Chen}, \bibinfo{person}{Zhaojun
  Jiang}, \bibinfo{person}{Duowei Xia}, \bibinfo{person}{Zebin Cai},
  \bibinfo{person}{Lingyun Sun}, \bibinfo{person}{Peter Childs}, {and}
  \bibinfo{person}{Haoyu Zuo}.} \bibinfo{year}{2024}\natexlab{a}.
\newblock \showarticletitle{BIDTrainer: An LLMs-driven Education Tool for
  Enhancing the Understanding and Reasoning in Bio-inspired Design}. In
  \bibinfo{booktitle}{\emph{Proceedings of the CHI Conference on Human Factors
  in Computing Systems}} (Honolulu, HI, USA) \emph{(\bibinfo{series}{CHI
  '24})}. \bibinfo{publisher}{Association for Computing Machinery},
  \bibinfo{address}{New York, NY, USA}, Article \bibinfo{articleno}{676},
  \bibinfo{numpages}{20}~pages.
\newblock
\showISBNx{9798400703300}
\urldef\tempurl%
\url{https://doi.org/10.1145/3613904.3642887}
\showDOI{\tempurl}


\bibitem[Chen et~al\mbox{.}(2024b)]%
        {chen_beyond_2024}
\bibfield{author}{\bibinfo{person}{Qing Chen}, \bibinfo{person}{Wei Shuai},
  \bibinfo{person}{Jiyao Zhang}, \bibinfo{person}{Zhida Sun}, {and}
  \bibinfo{person}{Nan Cao}.} \bibinfo{year}{2024}\natexlab{b}.
\newblock \showarticletitle{Beyond {Numbers}: {Creating} {Analogies} to
  {Enhance} {Data} {Comprehension} and {Communication} with {Generative} {AI}}.
  In \bibinfo{booktitle}{\emph{Proceedings of the {CHI} {Conference} on {Human}
  {Factors} in {Computing} {Systems}}} \emph{(\bibinfo{series}{{CHI} '24})}.
  \bibinfo{publisher}{Association for Computing Machinery},
  \bibinfo{address}{New York, NY, USA}, \bibinfo{pages}{1--14}.
\newblock
\showISBNx{9798400703300}
\urldef\tempurl%
\url{https://doi.org/10.1145/3613904.3642480}
\showDOI{\tempurl}


\bibitem[Chen et~al\mbox{.}(2024c)]%
        {chen2024stugptviz}
\bibfield{author}{\bibinfo{person}{Zixin Chen}, \bibinfo{person}{Jiachen Wang},
  \bibinfo{person}{Meng Xia}, \bibinfo{person}{Kento Shigyo},
  \bibinfo{person}{Dingdong Liu}, \bibinfo{person}{Rong Zhang}, {and}
  \bibinfo{person}{Huamin Qu}.} \bibinfo{year}{2024}\natexlab{c}.
\newblock \showarticletitle{StuGPTViz: A Visual Analytics Approach to
  Understand Student-ChatGPT Interactions}.
\newblock \bibinfo{journal}{\emph{arXiv preprint arXiv:2407.12423}}
  (\bibinfo{year}{2024}).
\newblock


\bibitem[Davies(1985)]%
        {davies_analogy_1985}
\bibfield{author}{\bibinfo{person}{Todd Davies}.}
  \bibinfo{year}{1985}\natexlab{}.
\newblock \showarticletitle{Analogy}.
\newblock \bibinfo{journal}{\emph{CSLI Informal Notes Series IN-CSLI-85-4,
  Center for the Study of Language and Information, Stanford}}
  (\bibinfo{year}{1985}).
\newblock


\bibitem[Devlin et~al\mbox{.}(2019a)]%
        {devlin-etal-2019-bert}
\bibfield{author}{\bibinfo{person}{Jacob Devlin}, \bibinfo{person}{Ming-Wei
  Chang}, \bibinfo{person}{Kenton Lee}, {and} \bibinfo{person}{Kristina
  Toutanova}.} \bibinfo{year}{2019}\natexlab{a}.
\newblock \showarticletitle{{BERT}: Pre-training of Deep Bidirectional
  Transformers for Language Understanding}. In
  \bibinfo{booktitle}{\emph{Proceedings of the 2019 Conference of the North
  {A}merican Chapter of the Association for Computational Linguistics: Human
  Language Technologies, Volume 1 (Long and Short Papers)}},
  \bibfield{editor}{\bibinfo{person}{Jill Burstein}, \bibinfo{person}{Christy
  Doran}, {and} \bibinfo{person}{Thamar Solorio}} (Eds.).
  \bibinfo{publisher}{Association for Computational Linguistics},
  \bibinfo{address}{Minneapolis, Minnesota}, \bibinfo{pages}{4171--4186}.
\newblock
\urldef\tempurl%
\url{https://doi.org/10.18653/v1/N19-1423}
\showDOI{\tempurl}


\bibitem[Devlin et~al\mbox{.}(2019b)]%
        {devlin_bert_2019}
\bibfield{author}{\bibinfo{person}{Jacob Devlin}, \bibinfo{person}{Ming-Wei
  Chang}, \bibinfo{person}{Kenton Lee}, {and} \bibinfo{person}{Kristina
  Toutanova}.} \bibinfo{year}{2019}\natexlab{b}.
\newblock \showarticletitle{{BERT}: {Pre}-training of {Deep} {Bidirectional}
  {Transformers} for {Language} {Understanding}}. In
  \bibinfo{booktitle}{\emph{Proceedings of the 2019 {Conference} of the {North}
  {American} {Chapter} of the {Association} for {Computational} {Linguistics}:
  {Human} {Language} {Technologies}, {Volume} 1 ({Long} and {Short}
  {Papers})}}. \bibinfo{publisher}{Association for Computational Linguistics},
  \bibinfo{address}{Minneapolis, Minnesota}, \bibinfo{pages}{4171--4186}.
\newblock
\urldef\tempurl%
\url{https://doi.org/10.18653/v1/N19-1423}
\showDOI{\tempurl}


\bibitem[Ding et~al\mbox{.}(2023)]%
        {ding_fluid_2023}
\bibfield{author}{\bibinfo{person}{Zijian Ding}, \bibinfo{person}{Arvind
  Srinivasan}, \bibinfo{person}{Stephen Macneil}, {and} \bibinfo{person}{Joel
  Chan}.} \bibinfo{year}{2023}\natexlab{}.
\newblock \showarticletitle{Fluid {Transformers} and {Creative} {Analogies}:
  {Exploring} {Large} {Language} {Models}’ {Capacity} for {Augmenting}
  {Cross}-{Domain} {Analogical} {Creativity}}. In
  \bibinfo{booktitle}{\emph{Proceedings of the 15th {Conference} on
  {Creativity} and {Cognition}}} \emph{(\bibinfo{series}{C\&amp;{C} '23})}.
  \bibinfo{publisher}{Association for Computing Machinery},
  \bibinfo{address}{New York, NY, USA}, \bibinfo{pages}{489--505}.
\newblock
\showISBNx{9798400701801}
\urldef\tempurl%
\url{https://doi.org/10.1145/3591196.3593516}
\showDOI{\tempurl}


\bibitem[Duit et~al\mbox{.}(1991)]%
        {duit1991role}
\bibfield{author}{\bibinfo{person}{Reinders Duit} {et~al\mbox{.}}}
  \bibinfo{year}{1991}\natexlab{}.
\newblock \showarticletitle{On the role of analogies and metaphors in learning
  science}.
\newblock \bibinfo{journal}{\emph{Science education}} \bibinfo{volume}{75},
  \bibinfo{number}{6} (\bibinfo{year}{1991}), \bibinfo{pages}{649--672}.
\newblock


\bibitem[Fan et~al\mbox{.}(2024)]%
        {lessonplanner2024uist}
\bibfield{author}{\bibinfo{person}{Haoxiang Fan}, \bibinfo{person}{Guanzheng
  Chen}, \bibinfo{person}{Xingbo Wang}, {and} \bibinfo{person}{Zhenhui Peng}.}
  \bibinfo{year}{2024}\natexlab{}.
\newblock \showarticletitle{LessonPlanner: Assisting Novice Teachers to Prepare
  Pedagogy-Driven Lesson Plans with Large Language Models}. In
  \bibinfo{booktitle}{\emph{Proceedings of the 37th Annual ACM Symposium on
  User Interface Software and Technology}} (Pittsburgh, PA, USA)
  \emph{(\bibinfo{series}{UIST '24})}. \bibinfo{publisher}{Association for
  Computing Machinery}, \bibinfo{address}{New York, NY, USA}, Article
  \bibinfo{articleno}{146}, \bibinfo{numpages}{20}~pages.
\newblock
\showISBNx{9798400706288}
\urldef\tempurl%
\url{https://doi.org/10.1145/3654777.3676390}
\showDOI{\tempurl}


\bibitem[Galesic and Garcia-Retamero(2013)]%
        {galesic_using_2013}
\bibfield{author}{\bibinfo{person}{Mirta Galesic} {and} \bibinfo{person}{Rocio
  Garcia-Retamero}.} \bibinfo{year}{2013}\natexlab{}.
\newblock \showarticletitle{Using {Analogies} to {Communicate} {Information}
  about {Health} {Risks}}.
\newblock \bibinfo{journal}{\emph{Applied Cognitive Psychology}}
  \bibinfo{volume}{27}, \bibinfo{number}{1} (\bibinfo{year}{2013}),
  \bibinfo{pages}{33--42}.
\newblock
\showISSN{1099-0720}
\urldef\tempurl%
\url{https://doi.org/10.1002/acp.2866}
\showDOI{\tempurl}
\newblock
\shownote{\_eprint: https://onlinelibrary.wiley.com/doi/pdf/10.1002/acp.2866}.


\bibitem[Gao et~al\mbox{.}(2024)]%
        {gao2024fine}
\bibfield{author}{\bibinfo{person}{Lin Gao}, \bibinfo{person}{Jing Lu},
  \bibinfo{person}{Zekai Shao}, \bibinfo{person}{Ziyue Lin},
  \bibinfo{person}{Shengbin Yue}, \bibinfo{person}{Chiokit Ieong},
  \bibinfo{person}{Yi Sun}, \bibinfo{person}{Rory~James Zauner},
  \bibinfo{person}{Zhongyu Wei}, {and} \bibinfo{person}{Siming Chen}.}
  \bibinfo{year}{2024}\natexlab{}.
\newblock \showarticletitle{Fine-Tuned Large Language Model for Visualization
  System: A Study on Self-Regulated Learning in Education}.
\newblock \bibinfo{journal}{\emph{IEEE Transactions on Visualization and
  Computer Graphics}} (\bibinfo{year}{2024}), \bibinfo{pages}{1--11}.
\newblock
\urldef\tempurl%
\url{https://doi.org/10.1109/TVCG.2024.3456145}
\showDOI{\tempurl}


\bibitem[Gentner(1983)]%
        {gentner1983structure}
\bibfield{author}{\bibinfo{person}{Dedre Gentner}.}
  \bibinfo{year}{1983}\natexlab{}.
\newblock \showarticletitle{Structure-mapping: A theoretical framework for
  analogy}.
\newblock \bibinfo{journal}{\emph{Cognitive science}} \bibinfo{volume}{7},
  \bibinfo{number}{2} (\bibinfo{year}{1983}), \bibinfo{pages}{155--170}.
\newblock


\bibitem[Gentner and Forbus(2011)]%
        {gentner_computational_2011}
\bibfield{author}{\bibinfo{person}{Dedre Gentner} {and}
  \bibinfo{person}{Kenneth~D Forbus}.} \bibinfo{year}{2011}\natexlab{}.
\newblock \showarticletitle{Computational models of analogy}.
\newblock \bibinfo{journal}{\emph{Wiley interdisciplinary reviews: cognitive
  science}} \bibinfo{volume}{2}, \bibinfo{number}{3} (\bibinfo{year}{2011}),
  \bibinfo{pages}{266--276}.
\newblock
\newblock
\shownote{Publisher: Wiley Online Library}.


\bibitem[Gentner and Hoyos(2017)]%
        {gentner2017analogy}
\bibfield{author}{\bibinfo{person}{Dedre Gentner} {and}
  \bibinfo{person}{Christian Hoyos}.} \bibinfo{year}{2017}\natexlab{}.
\newblock \showarticletitle{Analogy and abstraction}.
\newblock \bibinfo{journal}{\emph{Topics in cognitive science}}
  \bibinfo{volume}{9}, \bibinfo{number}{3} (\bibinfo{year}{2017}),
  \bibinfo{pages}{672--693}.
\newblock


\bibitem[Gentner and Markman(1997)]%
        {gentner_structure_1997}
\bibfield{author}{\bibinfo{person}{Dedre Gentner} {and}
  \bibinfo{person}{Arthur~B Markman}.} \bibinfo{year}{1997}\natexlab{}.
\newblock \showarticletitle{Structure mapping in analogy and similarity.}
\newblock \bibinfo{journal}{\emph{American psychologist}} \bibinfo{volume}{52},
  \bibinfo{number}{1} (\bibinfo{year}{1997}), \bibinfo{pages}{45}.
\newblock
\newblock
\shownote{Publisher: American Psychological Association}.


\bibitem[Gero et~al\mbox{.}(2024)]%
        {supporting2024gero}
\bibfield{author}{\bibinfo{person}{Katy~Ilonka Gero}, \bibinfo{person}{Chelse
  Swoopes}, \bibinfo{person}{Ziwei Gu}, \bibinfo{person}{Jonathan~K.
  Kummerfeld}, {and} \bibinfo{person}{Elena~L. Glassman}.}
  \bibinfo{year}{2024}\natexlab{}.
\newblock \showarticletitle{Supporting Sensemaking of Large Language Model
  Outputs at Scale}. In \bibinfo{booktitle}{\emph{Proceedings of the CHI
  Conference on Human Factors in Computing Systems}} (Honolulu, HI, USA)
  \emph{(\bibinfo{series}{CHI '24})}. \bibinfo{publisher}{Association for
  Computing Machinery}, \bibinfo{address}{New York, NY, USA}, Article
  \bibinfo{articleno}{838}, \bibinfo{numpages}{21}~pages.
\newblock
\showISBNx{9798400703300}
\urldef\tempurl%
\url{https://doi.org/10.1145/3613904.3642139}
\showDOI{\tempurl}


\bibitem[Gick and Holyoak(1980)]%
        {gick_analogical_1980}
\bibfield{author}{\bibinfo{person}{Mary~L. Gick} {and}
  \bibinfo{person}{Keith~J. Holyoak}.} \bibinfo{year}{1980}\natexlab{}.
\newblock \showarticletitle{Analogical problem solving}.
\newblock \bibinfo{journal}{\emph{Cognitive Psychology}} \bibinfo{volume}{12},
  \bibinfo{number}{3} (\bibinfo{date}{July} \bibinfo{year}{1980}),
  \bibinfo{pages}{306--355}.
\newblock
\showISSN{0010-0285}
\urldef\tempurl%
\url{https://doi.org/10.1016/0010-0285(80)90013-4}
\showDOI{\tempurl}


\bibitem[Gick and Holyoak(1983)]%
        {gick_schema_1983}
\bibfield{author}{\bibinfo{person}{Mary~L. Gick} {and}
  \bibinfo{person}{Keith~J. Holyoak}.} \bibinfo{year}{1983}\natexlab{}.
\newblock \showarticletitle{Schema induction and analogical transfer}.
\newblock \bibinfo{journal}{\emph{Cognitive Psychology}} \bibinfo{volume}{15},
  \bibinfo{number}{1} (\bibinfo{date}{Jan.} \bibinfo{year}{1983}),
  \bibinfo{pages}{1--38}.
\newblock
\showISSN{0010-0285}
\urldef\tempurl%
\url{https://doi.org/10.1016/0010-0285(83)90002-6}
\showDOI{\tempurl}


\bibitem[Gilon et~al\mbox{.}(2018)]%
        {analogymining2018Gilon}
\bibfield{author}{\bibinfo{person}{Karni Gilon}, \bibinfo{person}{Joel Chan},
  \bibinfo{person}{Felicia~Y. Ng}, \bibinfo{person}{Hila Liifshitz-Assaf},
  \bibinfo{person}{Aniket Kittur}, {and} \bibinfo{person}{Dafna Shahaf}.}
  \bibinfo{year}{2018}\natexlab{}.
\newblock \showarticletitle{Analogy Mining for Specific Design Needs}. In
  \bibinfo{booktitle}{\emph{Proceedings of the 2018 CHI Conference on Human
  Factors in Computing Systems}} (Montreal QC, Canada)
  \emph{(\bibinfo{series}{CHI '18})}. \bibinfo{publisher}{Association for
  Computing Machinery}, \bibinfo{address}{New York, NY, USA},
  \bibinfo{pages}{1–11}.
\newblock
\showISBNx{9781450356206}
\urldef\tempurl%
\url{https://doi.org/10.1145/3173574.3173695}
\showDOI{\tempurl}


\bibitem[Gladkova et~al\mbox{.}(2016)]%
        {gladkova_analogy-based_2016}
\bibfield{author}{\bibinfo{person}{Anna Gladkova}, \bibinfo{person}{Aleksandr
  Drozd}, {and} \bibinfo{person}{Satoshi Matsuoka}.}
  \bibinfo{year}{2016}\natexlab{}.
\newblock \showarticletitle{Analogy-based detection of morphological and
  semantic relations with word embeddings: what works and what doesn't.}. In
  \bibinfo{booktitle}{\emph{Proceedings of the {NAACL} {Student} {Research}
  {Workshop}}}. \bibinfo{publisher}{Association for Computational Linguistics},
  \bibinfo{address}{San Diego, California}, \bibinfo{pages}{8--15}.
\newblock
\urldef\tempurl%
\url{https://doi.org/10.18653/v1/N16-2002}
\showDOI{\tempurl}


\bibitem[Goel(1997)]%
        {goel_design_1997}
\bibfield{author}{\bibinfo{person}{Ashok~K Goel}.}
  \bibinfo{year}{1997}\natexlab{}.
\newblock \showarticletitle{Design, analogy, and creativity}.
\newblock \bibinfo{journal}{\emph{IEEE expert}} \bibinfo{volume}{12},
  \bibinfo{number}{3} (\bibinfo{year}{1997}), \bibinfo{pages}{62--70}.
\newblock
\newblock
\shownote{Publisher: IEEE}.


\bibitem[Gray and Holyoak(2021)]%
        {gray_teaching_2021}
\bibfield{author}{\bibinfo{person}{Maureen~E. Gray} {and}
  \bibinfo{person}{Keith~J. Holyoak}.} \bibinfo{year}{2021}\natexlab{}.
\newblock \showarticletitle{Teaching by {Analogy}: {From} {Theory} to
  {Practice}}.
\newblock \bibinfo{journal}{\emph{Mind, Brain, and Education}}
  \bibinfo{volume}{15}, \bibinfo{number}{3} (\bibinfo{year}{2021}),
  \bibinfo{pages}{250--263}.
\newblock
\showISSN{1751-228X}
\urldef\tempurl%
\url{https://doi.org/10.1111/mbe.12288}
\showDOI{\tempurl}
\newblock
\shownote{\_eprint: https://onlinelibrary.wiley.com/doi/pdf/10.1111/mbe.12288}.


\bibitem[Hesse(1959)]%
        {hesse1959defining}
\bibfield{author}{\bibinfo{person}{Mary~B Hesse}.}
  \bibinfo{year}{1959}\natexlab{}.
\newblock \showarticletitle{On defining analogy}. In
  \bibinfo{booktitle}{\emph{Proceedings of the Aristotelian Society}},
  Vol.~\bibinfo{volume}{60}. JSTOR, \bibinfo{pages}{79--100}.
\newblock


\bibitem[Hnatyshyn et~al\mbox{.}(2024)]%
        {capturing2024hnatyshyn}
\bibfield{author}{\bibinfo{person}{Rostyslav Hnatyshyn}, \bibinfo{person}{Jiayi
  Hong}, \bibinfo{person}{Ross Maciejewski}, \bibinfo{person}{Christopher
  Norby}, {and} \bibinfo{person}{Carlo~C. Maley}.}
  \bibinfo{year}{2024}\natexlab{}.
\newblock \showarticletitle{Capturing Cancer as Music: Cancer Mechanisms
  Expressed through Musification}. In \bibinfo{booktitle}{\emph{Proceedings of
  the CHI Conference on Human Factors in Computing Systems}} (Honolulu, HI,
  USA) \emph{(\bibinfo{series}{CHI '24})}. \bibinfo{publisher}{Association for
  Computing Machinery}, \bibinfo{address}{New York, NY, USA}, Article
  \bibinfo{articleno}{727}, \bibinfo{numpages}{11}~pages.
\newblock
\showISBNx{9798400703300}
\urldef\tempurl%
\url{https://doi.org/10.1145/3613904.3642153}
\showDOI{\tempurl}


\bibitem[Holyoak and Thagard(1996)]%
        {holyoak_mental_1996}
\bibfield{author}{\bibinfo{person}{Keith~J Holyoak} {and} \bibinfo{person}{Paul
  Thagard}.} \bibinfo{year}{1996}\natexlab{}.
\newblock \bibinfo{booktitle}{\emph{Mental leaps: {Analogy} in creative
  thought}}.
\newblock \bibinfo{publisher}{MIT press}.
\newblock


\bibitem[Hu et~al\mbox{.}(2023)]%
        {hu_-context_2023}
\bibfield{author}{\bibinfo{person}{Xiaoyang Hu}, \bibinfo{person}{Shane
  Storks}, \bibinfo{person}{Richard Lewis}, {and} \bibinfo{person}{Joyce
  Chai}.} \bibinfo{year}{2023}\natexlab{}.
\newblock \showarticletitle{In-{Context} {Analogical} {Reasoning} with
  {Pre}-{Trained} {Language} {Models}}. In
  \bibinfo{booktitle}{\emph{Proceedings of the 61st {Annual} {Meeting} of the
  {Association} for {Computational} {Linguistics} ({Volume} 1: {Long}
  {Papers})}}. \bibinfo{publisher}{Association for Computational Linguistics},
  \bibinfo{address}{Toronto, Canada}, \bibinfo{pages}{1953--1969}.
\newblock
\urldef\tempurl%
\url{https://doi.org/10.18653/v1/2023.acl-long.109}
\showDOI{\tempurl}


\bibitem[Hullman et~al\mbox{.}(2018)]%
        {improving2018hullman}
\bibfield{author}{\bibinfo{person}{Jessica Hullman}, \bibinfo{person}{Yea-Seul
  Kim}, \bibinfo{person}{Francis Nguyen}, \bibinfo{person}{Lauren Speers},
  {and} \bibinfo{person}{Maneesh Agrawala}.} \bibinfo{year}{2018}\natexlab{}.
\newblock \showarticletitle{Improving Comprehension of Measurements Using
  Concrete Re-expression Strategies}. In \bibinfo{booktitle}{\emph{Proceedings
  of the 2018 CHI Conference on Human Factors in Computing Systems}} (Montreal
  QC, Canada) \emph{(\bibinfo{series}{CHI '18})}.
  \bibinfo{publisher}{Association for Computing Machinery},
  \bibinfo{address}{New York, NY, USA}, \bibinfo{pages}{1–12}.
\newblock
\showISBNx{9781450356206}
\urldef\tempurl%
\url{https://doi.org/10.1145/3173574.3173608}
\showDOI{\tempurl}


\bibitem[Jiayang et~al\mbox{.}(2023)]%
        {jiayang_storyanalogy_2023}
\bibfield{author}{\bibinfo{person}{Cheng Jiayang}, \bibinfo{person}{Lin Qiu},
  \bibinfo{person}{Tsz Chan}, \bibinfo{person}{Tianqing Fang},
  \bibinfo{person}{Weiqi Wang}, \bibinfo{person}{Chunkit Chan},
  \bibinfo{person}{Dongyu Ru}, \bibinfo{person}{Qipeng Guo},
  \bibinfo{person}{Hongming Zhang}, \bibinfo{person}{Yangqiu Song},
  \bibinfo{person}{Yue Zhang}, {and} \bibinfo{person}{Zheng Zhang}.}
  \bibinfo{year}{2023}\natexlab{}.
\newblock \showarticletitle{{StoryAnalogy}: {Deriving} {Story}-level
  {Analogies} from {Large} {Language} {Models} to {Unlock} {Analogical}
  {Understanding}}. In \bibinfo{booktitle}{\emph{Proceedings of the 2023
  {Conference} on {Empirical} {Methods} in {Natural} {Language} {Processing}}},
  \bibfield{editor}{\bibinfo{person}{Houda Bouamor}, \bibinfo{person}{Juan
  Pino}, {and} \bibinfo{person}{Kalika Bali}} (Eds.).
  \bibinfo{publisher}{Association for Computational Linguistics},
  \bibinfo{address}{Singapore}, \bibinfo{pages}{11518--11537}.
\newblock
\urldef\tempurl%
\url{https://doi.org/10.18653/v1/2023.emnlp-main.706}
\showDOI{\tempurl}


\bibitem[Kang et~al\mbox{.}(2024)]%
        {kang_biospark_2024}
\bibfield{author}{\bibinfo{person}{Hyeonsu~B Kang}, \bibinfo{person}{David
  Chuan-En Lin}, \bibinfo{person}{Nikolas Martelaro}, \bibinfo{person}{Aniket
  Kittur}, \bibinfo{person}{Yan-Ying Chen}, {and} \bibinfo{person}{Matthew~K.
  Hong}.} \bibinfo{year}{2024}\natexlab{}.
\newblock \showarticletitle{{BioSpark}: {An} {End}-to-{End} {Generative}
  {System} for {Biological}-{Analogical} {Inspirations} and {Ideation}}. In
  \bibinfo{booktitle}{\emph{Extended {Abstracts} of the 2024 {CHI} {Conference}
  on {Human} {Factors} in {Computing} {Systems}}} \emph{(\bibinfo{series}{{CHI}
  {EA} '24})}. \bibinfo{publisher}{Association for Computing Machinery},
  \bibinfo{address}{New York, NY, USA}.
\newblock
\showISBNx{9798400703317}
\urldef\tempurl%
\url{https://doi.org/10.1145/3613905.3651035}
\showDOI{\tempurl}


\bibitem[Kang et~al\mbox{.}(2022)]%
        {kang_augmenting_2022}
\bibfield{author}{\bibinfo{person}{Hyeonsu~B. Kang}, \bibinfo{person}{Xin
  Qian}, \bibinfo{person}{Tom Hope}, \bibinfo{person}{Dafna Shahaf},
  \bibinfo{person}{Joel Chan}, {and} \bibinfo{person}{Aniket Kittur}.}
  \bibinfo{year}{2022}\natexlab{}.
\newblock \showarticletitle{Augmenting {Scientific} {Creativity} with an
  {Analogical} {Search} {Engine}}.
\newblock \bibinfo{journal}{\emph{ACM Transactions on Computer-Human
  Interaction}} \bibinfo{volume}{29}, \bibinfo{number}{6} (\bibinfo{date}{Nov.}
  \bibinfo{year}{2022}), \bibinfo{pages}{57:1--57:36}.
\newblock
\showISSN{1073-0516}
\urldef\tempurl%
\url{https://doi.org/10.1145/3530013}
\showDOI{\tempurl}


\bibitem[Kao(2020)]%
        {kao_how_2020}
\bibfield{author}{\bibinfo{person}{Chen-Yao Kao}.}
  \bibinfo{year}{2020}\natexlab{}.
\newblock \showarticletitle{How figurativity of analogy affects creativity:
  {The} application of four-term analogies to teaching for creativity}.
\newblock \bibinfo{journal}{\emph{Thinking skills and creativity}}
  \bibinfo{volume}{36} (\bibinfo{year}{2020}), \bibinfo{pages}{100653}.
\newblock
\newblock
\shownote{Publisher: Elsevier}.


\bibitem[Kasneci et~al\mbox{.}(2023)]%
        {kasneci2023chatgpt}
\bibfield{author}{\bibinfo{person}{Enkelejda Kasneci}, \bibinfo{person}{Kathrin
  Se{\ss}ler}, \bibinfo{person}{Stefan K{\"u}chemann}, \bibinfo{person}{Maria
  Bannert}, \bibinfo{person}{Daryna Dementieva}, \bibinfo{person}{Frank
  Fischer}, \bibinfo{person}{Urs Gasser}, \bibinfo{person}{Georg Groh},
  \bibinfo{person}{Stephan G{\"u}nnemann}, \bibinfo{person}{Eyke
  H{\"u}llermeier}, {et~al\mbox{.}}} \bibinfo{year}{2023}\natexlab{}.
\newblock \showarticletitle{ChatGPT for good? On opportunities and challenges
  of large language models for education}.
\newblock \bibinfo{journal}{\emph{Learning and individual differences}}
  \bibinfo{volume}{103} (\bibinfo{year}{2023}), \bibinfo{pages}{102274}.
\newblock


\bibitem[Kazemitabaar et~al\mbox{.}(2024)]%
        {kazemitabaar_codeaid_2024}
\bibfield{author}{\bibinfo{person}{Majeed Kazemitabaar},
  \bibinfo{person}{Runlong Ye}, \bibinfo{person}{Xiaoning Wang},
  \bibinfo{person}{Austin~Zachary Henley}, \bibinfo{person}{Paul Denny},
  \bibinfo{person}{Michelle Craig}, {and} \bibinfo{person}{Tovi Grossman}.}
  \bibinfo{year}{2024}\natexlab{}.
\newblock \showarticletitle{{CodeAid}: {Evaluating} a {Classroom} {Deployment}
  of an {LLM}-based {Programming} {Assistant} that {Balances} {Student} and
  {Educator} {Needs}}. In \bibinfo{booktitle}{\emph{Proceedings of the {CHI}
  {Conference} on {Human} {Factors} in {Computing} {Systems}}}
  \emph{(\bibinfo{series}{{CHI} '24})}. \bibinfo{publisher}{Association for
  Computing Machinery}, \bibinfo{address}{New York, NY, USA},
  \bibinfo{pages}{1--20}.
\newblock
\showISBNx{9798400703300}
\urldef\tempurl%
\url{https://doi.org/10.1145/3613904.3642773}
\showDOI{\tempurl}


\bibitem[Kim et~al\mbox{.}(2022)]%
        {putting2022kim}
\bibfield{author}{\bibinfo{person}{Yea-Seul Kim}, \bibinfo{person}{Jake~M
  Hofman}, {and} \bibinfo{person}{Daniel~G Goldstein}.}
  \bibinfo{year}{2022}\natexlab{}.
\newblock \showarticletitle{Putting scientific results in perspective:
  Improving the communication of standardized effect sizes}. In
  \bibinfo{booktitle}{\emph{Proceedings of the 2022 CHI Conference on Human
  Factors in Computing Systems}} (New Orleans, LA, USA)
  \emph{(\bibinfo{series}{CHI '22})}. \bibinfo{publisher}{Association for
  Computing Machinery}, \bibinfo{address}{New York, NY, USA}, Article
  \bibinfo{articleno}{625}, \bibinfo{numpages}{14}~pages.
\newblock
\showISBNx{9781450391573}
\urldef\tempurl%
\url{https://doi.org/10.1145/3491102.3502053}
\showDOI{\tempurl}


\bibitem[Kim et~al\mbox{.}(2016)]%
        {generating2016kim}
\bibfield{author}{\bibinfo{person}{Yea-Seul Kim}, \bibinfo{person}{Jessica
  Hullman}, {and} \bibinfo{person}{Maneesh Agrawala}.}
  \bibinfo{year}{2016}\natexlab{}.
\newblock \showarticletitle{Generating Personalized Spatial Analogies for
  Distances and Areas}. In \bibinfo{booktitle}{\emph{Proceedings of the 2016
  CHI Conference on Human Factors in Computing Systems}} (San Jose, California,
  USA) \emph{(\bibinfo{series}{CHI '16})}. \bibinfo{publisher}{Association for
  Computing Machinery}, \bibinfo{address}{New York, NY, USA},
  \bibinfo{pages}{38–48}.
\newblock
\showISBNx{9781450333627}
\urldef\tempurl%
\url{https://doi.org/10.1145/2858036.2858440}
\showDOI{\tempurl}


\bibitem[Kumar et~al\mbox{.}(2015)]%
        {kumar2015stickipedia}
\bibfield{author}{\bibinfo{person}{Varun Kumar}, \bibinfo{person}{Savita Bhat},
  {and} \bibinfo{person}{Niranjan Pedanekar}.} \bibinfo{year}{2015}\natexlab{}.
\newblock \showarticletitle{Stickipedia: A search engine and repository for
  explanatory analogies}. In \bibinfo{booktitle}{\emph{2015 IEEE 15th
  International Conference on Advanced Learning Technologies}}. IEEE,
  \bibinfo{pages}{280--284}.
\newblock


\bibitem[Li et~al\mbox{.}(2024)]%
        {GeneratingContextualized}
\bibfield{author}{\bibinfo{person}{Ruijia Li}, \bibinfo{person}{Yiting Wang},
  \bibinfo{person}{Chanjin Zheng}, \bibinfo{person}{Yuan-Hao Jiang}, {and}
  \bibinfo{person}{Bo Jiang}.} \bibinfo{year}{2024}\natexlab{}.
\newblock \showarticletitle{Generating Contextualized Mathematics
  Multiple-Choice Questions Utilizing Large Language Models}. In
  \bibinfo{booktitle}{\emph{Artificial Intelligence in Education. Posters and
  Late Breaking Results, Workshops and Tutorials, Industry and Innovation
  Tracks, Practitioners, Doctoral Consortium and Blue Sky}},
  \bibfield{editor}{\bibinfo{person}{Andrew~M. Olney},
  \bibinfo{person}{Irene-Angelica Chounta}, \bibinfo{person}{Zitao Liu},
  \bibinfo{person}{Olga~C. Santos}, {and} \bibinfo{person}{Ig~Ibert
  Bittencourt}} (Eds.). \bibinfo{publisher}{Springer Nature Switzerland},
  \bibinfo{address}{Cham}, \bibinfo{pages}{494--501}.
\newblock
\showISBNx{978-3-031-64315-6}


\bibitem[Liu et~al\mbox{.}(2024)]%
        {liu2024large}
\bibfield{author}{\bibinfo{person}{Dancheng Liu}, \bibinfo{person}{Amir
  Nassereldine}, \bibinfo{person}{Ziming Yang}, \bibinfo{person}{Chenhui Xu},
  \bibinfo{person}{Yuting Hu}, \bibinfo{person}{Jiajie Li},
  \bibinfo{person}{Utkarsh Kumar}, \bibinfo{person}{Changjae Lee}, {and}
  \bibinfo{person}{Jinjun Xiong}.} \bibinfo{year}{2024}\natexlab{}.
\newblock \showarticletitle{Large Language Models have Intrinsic
  Self-Correction Ability}.
\newblock \bibinfo{journal}{\emph{arXiv preprint arXiv:2406.15673}}
  (\bibinfo{year}{2024}).
\newblock


\bibitem[Lu et~al\mbox{.}(2023)]%
        {readingquizmaker}
\bibfield{author}{\bibinfo{person}{Xinyi Lu}, \bibinfo{person}{Simin Fan},
  \bibinfo{person}{Jessica Houghton}, \bibinfo{person}{Lu Wang}, {and}
  \bibinfo{person}{Xu Wang}.} \bibinfo{year}{2023}\natexlab{}.
\newblock \showarticletitle{ReadingQuizMaker: A Human-NLP Collaborative System
  that Supports Instructors to Design High-Quality Reading Quiz Questions}. In
  \bibinfo{booktitle}{\emph{Proceedings of the 2023 CHI Conference on Human
  Factors in Computing Systems}} (Hamburg, Germany) \emph{(\bibinfo{series}{CHI
  '23})}. \bibinfo{publisher}{Association for Computing Machinery},
  \bibinfo{address}{New York, NY, USA}, Article \bibinfo{articleno}{454},
  \bibinfo{numpages}{18}~pages.
\newblock
\showISBNx{9781450394215}
\urldef\tempurl%
\url{https://doi.org/10.1145/3544548.3580957}
\showDOI{\tempurl}


\bibitem[Lyu et~al\mbox{.}(2024)]%
        {Lyu2024evaluating}
\bibfield{author}{\bibinfo{person}{Wenhan Lyu}, \bibinfo{person}{Yimeng Wang},
  \bibinfo{person}{Tingting~(Rachel) Chung}, \bibinfo{person}{Yifan Sun}, {and}
  \bibinfo{person}{Yixuan Zhang}.} \bibinfo{year}{2024}\natexlab{}.
\newblock \showarticletitle{Evaluating the Effectiveness of LLMs in
  Introductory Computer Science Education: A Semester-Long Field Study}. In
  \bibinfo{booktitle}{\emph{Proceedings of the Eleventh ACM Conference on
  Learning @ Scale}} (Atlanta, GA, USA) \emph{(\bibinfo{series}{L@S '24})}.
  \bibinfo{publisher}{Association for Computing Machinery},
  \bibinfo{address}{New York, NY, USA}, \bibinfo{pages}{63–74}.
\newblock
\showISBNx{9798400706332}
\urldef\tempurl%
\url{https://doi.org/10.1145/3657604.3662036}
\showDOI{\tempurl}


\bibitem[Mikolov et~al\mbox{.}(2013)]%
        {mikolov_linguistic_2013}
\bibfield{author}{\bibinfo{person}{Tomas Mikolov}, \bibinfo{person}{Wen-tau
  Yih}, {and} \bibinfo{person}{Geoffrey Zweig}.}
  \bibinfo{year}{2013}\natexlab{}.
\newblock \showarticletitle{Linguistic {Regularities} in {Continuous} {Space}
  {Word} {Representations}}. In \bibinfo{booktitle}{\emph{Proceedings of the
  2013 {Conference} of the {North} {American} {Chapter} of the {Association}
  for {Computational} {Linguistics}: {Human} {Language} {Technologies}}}.
  \bibinfo{publisher}{Association for Computational Linguistics},
  \bibinfo{address}{Atlanta, Georgia}, \bibinfo{pages}{746--751}.
\newblock
\urldef\tempurl%
\url{https://aclanthology.org/N13-1090}
\showURL{%
\tempurl}


\bibitem[Mitchell(2021)]%
        {mitchell_abstraction_2021}
\bibfield{author}{\bibinfo{person}{Melanie Mitchell}.}
  \bibinfo{year}{2021}\natexlab{}.
\newblock \showarticletitle{Abstraction and analogy-making in artificial
  intelligence}.
\newblock \bibinfo{journal}{\emph{Annals of the New York Academy of Sciences}}
  \bibinfo{volume}{1505}, \bibinfo{number}{1} (\bibinfo{year}{2021}),
  \bibinfo{pages}{79--101}.
\newblock
\newblock
\shownote{Publisher: Wiley Online Library}.


\bibitem[Ngoon et~al\mbox{.}(2024)]%
        {ngoon_classinsight_2024}
\bibfield{author}{\bibinfo{person}{Tricia~J. Ngoon}, \bibinfo{person}{S
  Sushil}, \bibinfo{person}{Angela~E.B. Stewart}, \bibinfo{person}{Ung-Sang
  Lee}, \bibinfo{person}{Saranya Venkatraman}, \bibinfo{person}{Neil Thawani},
  \bibinfo{person}{Prasenjit Mitra}, \bibinfo{person}{Sherice Clarke},
  \bibinfo{person}{John Zimmerman}, {and} \bibinfo{person}{Amy Ogan}.}
  \bibinfo{year}{2024}\natexlab{}.
\newblock \showarticletitle{{ClassInSight}: {Designing} {Conversation}
  {Support} {Tools} to {Visualize} {Classroom} {Discussion} for {Personalized}
  {Teacher} {Professional} {Development}}. In
  \bibinfo{booktitle}{\emph{Proceedings of the {CHI} {Conference} on {Human}
  {Factors} in {Computing} {Systems}}} \emph{(\bibinfo{series}{{CHI} '24})}.
  \bibinfo{publisher}{Association for Computing Machinery},
  \bibinfo{address}{New York, NY, USA}, \bibinfo{pages}{1--15}.
\newblock
\showISBNx{9798400703300}
\urldef\tempurl%
\url{https://doi.org/10.1145/3613904.3642487}
\showDOI{\tempurl}


\bibitem[Ni et~al\mbox{.}(2024)]%
        {edgeworth}
\bibfield{author}{\bibinfo{person}{Wode Ni}, \bibinfo{person}{Sam Estep},
  \bibinfo{person}{Hwei-Shin Harriman}, \bibinfo{person}{Kenneth~R. Koedinger},
  {and} \bibinfo{person}{Joshua Sunshine}.} \bibinfo{year}{2024}\natexlab{}.
\newblock \showarticletitle{Edgeworth: Efficient and Scalable Authoring of
  Visual Thinking Activities}. In \bibinfo{booktitle}{\emph{Proceedings of the
  Eleventh ACM Conference on Learning @ Scale}} (Atlanta, GA, USA)
  \emph{(\bibinfo{series}{L@S '24})}. \bibinfo{publisher}{Association for
  Computing Machinery}, \bibinfo{address}{New York, NY, USA},
  \bibinfo{pages}{98–109}.
\newblock
\showISBNx{9798400706332}
\urldef\tempurl%
\url{https://doi.org/10.1145/3657604.3662034}
\showDOI{\tempurl}


\bibitem[Oliva et~al\mbox{.}(2007)]%
        {oliva_teaching_2007}
\bibfield{author}{\bibinfo{person}{José Oliva}, \bibinfo{person}{Pilar
  Azcárate}, {and} \bibinfo{person}{Antonio Navarrete~Salvador}.}
  \bibinfo{year}{2007}\natexlab{}.
\newblock \showarticletitle{Teaching {Models} in the {Use} of {Analogies} as a
  {Resource} in the {Science} {Classroom}}.
\newblock \bibinfo{journal}{\emph{International Journal of Science Education -
  INT J SCI EDUC}}  \bibinfo{volume}{29} (\bibinfo{date}{Jan.}
  \bibinfo{year}{2007}), \bibinfo{pages}{45--66}.
\newblock
\urldef\tempurl%
\url{https://doi.org/10.1080/09500690600708444}
\showDOI{\tempurl}


\bibitem[{OpenAI}(2023)]%
        {openai_gpt-4_2023}
\bibfield{author}{\bibinfo{person}{{OpenAI}}.} \bibinfo{year}{2023}\natexlab{}.
\newblock \bibinfo{title}{{GPT}-4 {Technical} {Report}}.
\newblock
\newblock
\newblock
\shownote{\_eprint: 2303.08774}.


\bibitem[{OpenAI}(2024)]%
        {openai_o1_preview_2024}
\bibfield{author}{\bibinfo{person}{{OpenAI}}.} \bibinfo{year}{2024}\natexlab{}.
\newblock \bibinfo{title}{{Introducing OpenAI o1-preview}}.
\newblock
\newblock
\urldef\tempurl%
\url{https://openai.com/index/introducing-openai-o1-preview/}
\showURL{%
\tempurl}


\bibitem[Ouyang et~al\mbox{.}(2022)]%
        {ouyang_training_2022}
\bibfield{author}{\bibinfo{person}{Long Ouyang}, \bibinfo{person}{Jeffrey Wu},
  \bibinfo{person}{Xu Jiang}, \bibinfo{person}{Diogo Almeida},
  \bibinfo{person}{Carroll Wainwright}, \bibinfo{person}{Pamela Mishkin},
  \bibinfo{person}{Chong Zhang}, \bibinfo{person}{Sandhini Agarwal},
  \bibinfo{person}{Katarina Slama}, \bibinfo{person}{Alex Gray},
  \bibinfo{person}{John Schulman}, \bibinfo{person}{Jacob Hilton},
  \bibinfo{person}{Fraser Kelton}, \bibinfo{person}{Luke Miller},
  \bibinfo{person}{Maddie Simens}, \bibinfo{person}{Amanda Askell},
  \bibinfo{person}{Peter Welinder}, \bibinfo{person}{Paul Christiano},
  \bibinfo{person}{Jan Leike}, {and} \bibinfo{person}{Ryan Lowe}.}
  \bibinfo{year}{2022}\natexlab{}.
\newblock \showarticletitle{Training language models to follow instructions
  with human feedback}. In \bibinfo{booktitle}{\emph{Advances in {Neural}
  {Information} {Processing} {Systems}}},
  \bibfield{editor}{\bibinfo{person}{Alice~H. Oh}, \bibinfo{person}{Alekh
  Agarwal}, \bibinfo{person}{Danielle Belgrave}, {and}
  \bibinfo{person}{Kyunghyun Cho}} (Eds.).
\newblock
\urldef\tempurl%
\url{https://openreview.net/forum?id=TG8KACxEON}
\showURL{%
\tempurl}


\bibitem[P\"{a}\"{a}kk\"{o}nen et~al\mbox{.}(2020)]%
        {bureaucracy2020Pääkkönen}
\bibfield{author}{\bibinfo{person}{Juho P\"{a}\"{a}kk\"{o}nen},
  \bibinfo{person}{Matti Nelimarkka}, \bibinfo{person}{Jesse Haapoja}, {and}
  \bibinfo{person}{Airi Lampinen}.} \bibinfo{year}{2020}\natexlab{}.
\newblock \showarticletitle{Bureaucracy as a Lens for Analyzing and Designing
  Algorithmic Systems}. In \bibinfo{booktitle}{\emph{Proceedings of the 2020
  CHI Conference on Human Factors in Computing Systems}} (Honolulu, HI, USA)
  \emph{(\bibinfo{series}{CHI '20})}. \bibinfo{publisher}{Association for
  Computing Machinery}, \bibinfo{address}{New York, NY, USA},
  \bibinfo{pages}{1–14}.
\newblock
\showISBNx{9781450367080}
\urldef\tempurl%
\url{https://doi.org/10.1145/3313831.3376780}
\showDOI{\tempurl}


\bibitem[Pan et~al\mbox{.}(2023)]%
        {pan2023automatically}
\bibfield{author}{\bibinfo{person}{Liangming Pan}, \bibinfo{person}{Michael
  Saxon}, \bibinfo{person}{Wenda Xu}, \bibinfo{person}{Deepak Nathani},
  \bibinfo{person}{Xinyi Wang}, {and} \bibinfo{person}{William~Yang Wang}.}
  \bibinfo{year}{2023}\natexlab{}.
\newblock \showarticletitle{Automatically correcting large language models:
  Surveying the landscape of diverse self-correction strategies}.
\newblock \bibinfo{journal}{\emph{arXiv preprint arXiv:2308.03188}}
  (\bibinfo{year}{2023}).
\newblock


\bibitem[Radford et~al\mbox{.}(2019b)]%
        {radford2019language}
\bibfield{author}{\bibinfo{person}{Alec Radford}, \bibinfo{person}{Jeffrey Wu},
  \bibinfo{person}{Rewon Child}, \bibinfo{person}{David Luan},
  \bibinfo{person}{Dario Amodei}, \bibinfo{person}{Ilya Sutskever},
  {et~al\mbox{.}}} \bibinfo{year}{2019}\natexlab{b}.
\newblock \showarticletitle{Language models are unsupervised multitask
  learners}.
\newblock \bibinfo{journal}{\emph{OpenAI blog}} \bibinfo{volume}{1},
  \bibinfo{number}{8} (\bibinfo{year}{2019}), \bibinfo{pages}{9}.
\newblock


\bibitem[Radford et~al\mbox{.}(2019a)]%
        {radford_language_2019}
\bibfield{author}{\bibinfo{person}{Alec Radford}, \bibinfo{person}{Jeffrey Wu},
  \bibinfo{person}{Rewon Child}, \bibinfo{person}{David Luan},
  \bibinfo{person}{Dario Amodei}, \bibinfo{person}{Ilya Sutskever}, {and}
  \bibinfo{person}{{others}}.} \bibinfo{year}{2019}\natexlab{a}.
\newblock \showarticletitle{Language models are unsupervised multitask
  learners}.
\newblock \bibinfo{journal}{\emph{OpenAI blog}} \bibinfo{volume}{1},
  \bibinfo{number}{8} (\bibinfo{year}{2019}), \bibinfo{pages}{9}.
\newblock


\bibitem[Richland et~al\mbox{.}(2004)]%
        {richland_analogy_2004}
\bibfield{author}{\bibinfo{person}{Lindsey Richland}, \bibinfo{person}{Keith
  Holyoak}, {and} \bibinfo{person}{James Stigler}.}
  \bibinfo{year}{2004}\natexlab{}.
\newblock \showarticletitle{Analogy {Use} in {Eighth}-{Grade} {Mathematics}
  {Classrooms}}.
\newblock \bibinfo{journal}{\emph{Cognition and Instruction}}
  \bibinfo{volume}{22} (\bibinfo{date}{March} \bibinfo{year}{2004}),
  \bibinfo{pages}{37--60}.
\newblock
\urldef\tempurl%
\url{https://doi.org/10.1207/s1532690Xci2201_2}
\showDOI{\tempurl}


\bibitem[Richland and Simms(2015)]%
        {richland_analogy_2015}
\bibfield{author}{\bibinfo{person}{Lindsey~Engle Richland} {and}
  \bibinfo{person}{Nina Simms}.} \bibinfo{year}{2015}\natexlab{}.
\newblock \showarticletitle{Analogy, higher order thinking, and education}.
\newblock \bibinfo{journal}{\emph{WIREs Cognitive Science}}
  \bibinfo{volume}{6}, \bibinfo{number}{2} (\bibinfo{year}{2015}),
  \bibinfo{pages}{177--192}.
\newblock
\showISSN{1939-5086}
\urldef\tempurl%
\url{https://doi.org/10.1002/wcs.1336}
\showDOI{\tempurl}
\newblock
\shownote{\_eprint: https://onlinelibrary.wiley.com/doi/pdf/10.1002/wcs.1336}.


\bibitem[Richland et~al\mbox{.}(2007)]%
        {richland_cognitive_2007}
\bibfield{author}{\bibinfo{person}{Lindsey~E. Richland}, \bibinfo{person}{Osnat
  Zur}, {and} \bibinfo{person}{Keith~J. Holyoak}.}
  \bibinfo{year}{2007}\natexlab{}.
\newblock \showarticletitle{Cognitive {Supports} for {Analogies} in the
  {Mathematics} {Classroom}}.
\newblock \bibinfo{journal}{\emph{Science}} \bibinfo{volume}{316},
  \bibinfo{number}{5828} (\bibinfo{date}{May} \bibinfo{year}{2007}),
  \bibinfo{pages}{1128--1129}.
\newblock
\urldef\tempurl%
\url{https://doi.org/10.1126/science.1142103}
\showDOI{\tempurl}
\newblock
\shownote{Publisher: American Association for the Advancement of Science}.


\bibitem[Riederer et~al\mbox{.}(2018)]%
        {toput2018riederer}
\bibfield{author}{\bibinfo{person}{Christopher Riederer},
  \bibinfo{person}{Jake~M. Hofman}, {and} \bibinfo{person}{Daniel~G.
  Goldstein}.} \bibinfo{year}{2018}\natexlab{}.
\newblock \showarticletitle{To Put That in Perspective: Generating Analogies
  that Make Numbers Easier to Understand}. In
  \bibinfo{booktitle}{\emph{Proceedings of the 2018 CHI Conference on Human
  Factors in Computing Systems}} (Montreal QC, Canada)
  \emph{(\bibinfo{series}{CHI '18})}. \bibinfo{publisher}{Association for
  Computing Machinery}, \bibinfo{address}{New York, NY, USA},
  \bibinfo{pages}{1–10}.
\newblock
\showISBNx{9781450356206}
\urldef\tempurl%
\url{https://doi.org/10.1145/3173574.3174122}
\showDOI{\tempurl}


\bibitem[Shi et~al\mbox{.}(2025)]%
        {SHI2025125723}
\bibfield{author}{\bibinfo{person}{Jinxin Shi}, \bibinfo{person}{Jiabao Zhao},
  \bibinfo{person}{Xingjiao Wu}, \bibinfo{person}{Ruyi Xu},
  \bibinfo{person}{Yuan-Hao Jiang}, {and} \bibinfo{person}{Liang He}.}
  \bibinfo{year}{2025}\natexlab{}.
\newblock \showarticletitle{Mitigating reasoning hallucination through
  Multi-agent Collaborative Filtering}.
\newblock \bibinfo{journal}{\emph{Expert Systems with Applications}}
  \bibinfo{volume}{263} (\bibinfo{year}{2025}), \bibinfo{pages}{125723}.
\newblock
\showISSN{0957-4174}
\urldef\tempurl%
\url{https://doi.org/10.1016/j.eswa.2024.125723}
\showDOI{\tempurl}


\bibitem[Spatharioti et~al\mbox{.}(2024)]%
        {spatharioti_using_2024}
\bibfield{author}{\bibinfo{person}{Sofia~Eleni Spatharioti},
  \bibinfo{person}{Daniel~G Goldstein}, {and} \bibinfo{person}{Jake~M Hofman}.}
  \bibinfo{year}{2024}\natexlab{}.
\newblock \showarticletitle{Using {Open} {Data} to {Automatically} {Generate}
  {Localized} {Analogies}}. In \bibinfo{booktitle}{\emph{Proceedings of the
  {CHI} {Conference} on {Human} {Factors} in {Computing} {Systems}}}
  \emph{(\bibinfo{series}{{CHI} '24})}. \bibinfo{publisher}{Association for
  Computing Machinery}, \bibinfo{address}{New York, NY, USA}.
\newblock
\showISBNx{9798400703300}
\urldef\tempurl%
\url{https://doi.org/10.1145/3613904.3642638}
\showDOI{\tempurl}
\newblock
\shownote{event-place: Honolulu, HI, USA}.


\bibitem[Sultan et~al\mbox{.}(2024)]%
        {sultan_parallelparc_2024}
\bibfield{author}{\bibinfo{person}{Oren Sultan}, \bibinfo{person}{Yonatan
  Bitton}, \bibinfo{person}{Ron Yosef}, {and} \bibinfo{person}{Dafna Shahaf}.}
  \bibinfo{year}{2024}\natexlab{}.
\newblock \showarticletitle{{ParallelPARC}: {A} {Scalable} {Pipeline} for
  {Generating} {Natural}-{Language} {Analogies}}.
\newblock \bibinfo{journal}{\emph{arXiv preprint arXiv:2403.01139}}
  (\bibinfo{year}{2024}).
\newblock


\bibitem[Sultan and Shahaf(2022)]%
        {sultan_life_2022}
\bibfield{author}{\bibinfo{person}{Oren Sultan} {and} \bibinfo{person}{Dafna
  Shahaf}.} \bibinfo{year}{2022}\natexlab{}.
\newblock \showarticletitle{Life is a {Circus} and {We} are the {Clowns}:
  {Automatically} {Finding} {Analogies} between {Situations} and {Processes}}.
  In \bibinfo{booktitle}{\emph{Proceedings of the 2022 {Conference} on
  {Empirical} {Methods} in {Natural} {Language} {Processing}}}.
  \bibinfo{publisher}{Association for Computational Linguistics},
  \bibinfo{address}{Abu Dhabi, United Arab Emirates},
  \bibinfo{pages}{3547--3562}.
\newblock
\urldef\tempurl%
\url{https://aclanthology.org/2022.emnlp-main.232}
\showURL{%
\tempurl}


\bibitem[Tan and Subramonyam(2024)]%
        {tan_more_2024}
\bibfield{author}{\bibinfo{person}{Mei Tan} {and} \bibinfo{person}{Hari
  Subramonyam}.} \bibinfo{year}{2024}\natexlab{}.
\newblock \showarticletitle{More than {Model} {Documentation}: {Uncovering}
  {Teachers}' {Bespoke} {Information} {Needs} for {Informed} {Classroom}
  {Integration} of {ChatGPT}}. In \bibinfo{booktitle}{\emph{Proceedings of the
  {CHI} {Conference} on {Human} {Factors} in {Computing} {Systems}}}
  \emph{(\bibinfo{series}{{CHI} '24})}. \bibinfo{publisher}{Association for
  Computing Machinery}, \bibinfo{address}{New York, NY, USA},
  \bibinfo{pages}{1--19}.
\newblock
\showISBNx{9798400703300}
\urldef\tempurl%
\url{https://doi.org/10.1145/3613904.3642592}
\showDOI{\tempurl}


\bibitem[Tang et~al\mbox{.}(2024)]%
        {vizgroup}
\bibfield{author}{\bibinfo{person}{Xiaohang Tang}, \bibinfo{person}{Sam Wong},
  \bibinfo{person}{Kevin Pu}, \bibinfo{person}{Xi Chen},
  \bibinfo{person}{Yalong Yang}, {and} \bibinfo{person}{Yan Chen}.}
  \bibinfo{year}{2024}\natexlab{}.
\newblock \showarticletitle{VizGroup: An AI-assisted Event-driven System for
  Collaborative Programming Learning Analytics}. In
  \bibinfo{booktitle}{\emph{Proceedings of the 37th Annual ACM Symposium on
  User Interface Software and Technology}} (Pittsburgh, PA, USA)
  \emph{(\bibinfo{series}{UIST '24})}. \bibinfo{publisher}{Association for
  Computing Machinery}, \bibinfo{address}{New York, NY, USA}, Article
  \bibinfo{articleno}{93}, \bibinfo{numpages}{22}~pages.
\newblock
\showISBNx{9798400706288}
\urldef\tempurl%
\url{https://doi.org/10.1145/3654777.3676347}
\showDOI{\tempurl}


\bibitem[Team(2023)]%
        {team_gemini_2023}
\bibfield{author}{\bibinfo{person}{Gemini Team}.}
  \bibinfo{year}{2023}\natexlab{}.
\newblock \bibinfo{title}{Gemini: {A} {Family} of {Highly} {Capable}
  {Multimodal} {Models}}.
\newblock
\newblock
\newblock
\shownote{\_eprint: 2312.11805}.


\bibitem[Thagard(1992)]%
        {thagard_analogy_1992}
\bibfield{author}{\bibinfo{person}{Paul Thagard}.}
  \bibinfo{year}{1992}\natexlab{}.
\newblock \showarticletitle{Analogy, explanation, and education}.
\newblock \bibinfo{journal}{\emph{Journal of research in science teaching}}
  \bibinfo{volume}{29}, \bibinfo{number}{6} (\bibinfo{year}{1992}),
  \bibinfo{pages}{537--544}.
\newblock
\newblock
\shownote{Publisher: Wiley Online Library}.


\bibitem[Touvron et~al\mbox{.}(2023)]%
        {touvron_llama_2023}
\bibfield{author}{\bibinfo{person}{Hugo Touvron}, \bibinfo{person}{Thibaut
  Lavril}, \bibinfo{person}{Gautier Izacard}, \bibinfo{person}{Xavier
  Martinet}, \bibinfo{person}{Marie-Anne Lachaux}, \bibinfo{person}{Timothée
  Lacroix}, \bibinfo{person}{Baptiste Rozière}, \bibinfo{person}{Naman Goyal},
  \bibinfo{person}{Eric Hambro}, \bibinfo{person}{Faisal Azhar}, {and}
  \bibinfo{person}{{others}}.} \bibinfo{year}{2023}\natexlab{}.
\newblock \showarticletitle{Llama: {Open} and efficient foundation language
  models}.
\newblock \bibinfo{journal}{\emph{arXiv preprint arXiv:2302.13971}}
  (\bibinfo{year}{2023}).
\newblock


\bibitem[Treagust et~al\mbox{.}(1992)]%
        {treagust_science_1992}
\bibfield{author}{\bibinfo{person}{David Treagust}, \bibinfo{person}{Reinders
  Duit}, \bibinfo{person}{Paul Joslin}, {and} \bibinfo{person}{Ivo Lindauer}.}
  \bibinfo{year}{1992}\natexlab{}.
\newblock \showarticletitle{Science teachers' use of analogies: {Observations}
  from classroom practice}.
\newblock \bibinfo{journal}{\emph{International Journal of Science Education -
  INT J SCI EDUC}}  \bibinfo{volume}{14} (\bibinfo{date}{Oct.}
  \bibinfo{year}{1992}), \bibinfo{pages}{413--422}.
\newblock
\urldef\tempurl%
\url{https://doi.org/10.1080/0950069920140404}
\showDOI{\tempurl}


\bibitem[Turney et~al\mbox{.}(2003)]%
        {turney_combining_2003}
\bibfield{author}{\bibinfo{person}{Peter~D Turney}, \bibinfo{person}{Michael~L
  Littman}, \bibinfo{person}{Jeffrey Bigham}, {and} \bibinfo{person}{Victor
  Shnayder}.} \bibinfo{year}{2003}\natexlab{}.
\newblock \showarticletitle{Combining independent modules in lexical
  multiple-choice problems}.
\newblock \bibinfo{journal}{\emph{Recent Advances in Natural Language
  Processing III: Selected Papers from RANLP}}  \bibinfo{volume}{2003}
  (\bibinfo{year}{2003}), \bibinfo{pages}{101--110}.
\newblock


\bibitem[Vendetti et~al\mbox{.}(2015)]%
        {vendetti_analogical_2015}
\bibfield{author}{\bibinfo{person}{Michael~S. Vendetti},
  \bibinfo{person}{Bryan~J. Matlen}, \bibinfo{person}{Lindsey~E. Richland},
  {and} \bibinfo{person}{Silvia~A. Bunge}.} \bibinfo{year}{2015}\natexlab{}.
\newblock \showarticletitle{Analogical {Reasoning} in the {Classroom}:
  {Insights} {From} {Cognitive} {Science}}.
\newblock \bibinfo{journal}{\emph{Mind, Brain, and Education}}
  \bibinfo{volume}{9}, \bibinfo{number}{2} (\bibinfo{year}{2015}),
  \bibinfo{pages}{100--106}.
\newblock
\showISSN{1751-228X}
\urldef\tempurl%
\url{https://doi.org/10.1111/mbe.12080}
\showDOI{\tempurl}
\newblock
\shownote{\_eprint: https://onlinelibrary.wiley.com/doi/pdf/10.1111/mbe.12080}.


\bibitem[Wang et~al\mbox{.}(2024)]%
        {reelframer2024wang}
\bibfield{author}{\bibinfo{person}{Sitong Wang}, \bibinfo{person}{Samia Menon},
  \bibinfo{person}{Tao Long}, \bibinfo{person}{Keren Henderson},
  \bibinfo{person}{Dingzeyu Li}, \bibinfo{person}{Kevin Crowston},
  \bibinfo{person}{Mark Hansen}, \bibinfo{person}{Jeffrey~V Nickerson}, {and}
  \bibinfo{person}{Lydia~B Chilton}.} \bibinfo{year}{2024}\natexlab{}.
\newblock \showarticletitle{ReelFramer: Human-AI Co-Creation for News-to-Video
  Translation}. In \bibinfo{booktitle}{\emph{Proceedings of the CHI Conference
  on Human Factors in Computing Systems}} (Honolulu, HI, USA)
  \emph{(\bibinfo{series}{CHI '24})}. \bibinfo{publisher}{Association for
  Computing Machinery}, \bibinfo{address}{New York, NY, USA}, Article
  \bibinfo{articleno}{169}, \bibinfo{numpages}{20}~pages.
\newblock
\showISBNx{9798400703300}
\urldef\tempurl%
\url{https://doi.org/10.1145/3613904.3642868}
\showDOI{\tempurl}


\bibitem[Webb et~al\mbox{.}(2022)]%
        {webb_emergent_2022}
\bibfield{author}{\bibinfo{person}{Taylor Webb}, \bibinfo{person}{Keith~J
  Holyoak}, {and} \bibinfo{person}{Hongjing Lu}.}
  \bibinfo{year}{2022}\natexlab{}.
\newblock \showarticletitle{Emergent {Analogical} {Reasoning} in {Large}
  {Language} {Models}}.
\newblock \bibinfo{journal}{\emph{arXiv preprint arXiv:2212.09196}}
  (\bibinfo{year}{2022}).
\newblock


\bibitem[Wijesiriwardene et~al\mbox{.}(2023)]%
        {wijesiriwardene_analogical_2023}
\bibfield{author}{\bibinfo{person}{Thilini Wijesiriwardene},
  \bibinfo{person}{Ruwan Wickramarachchi}, \bibinfo{person}{Bimal Gajera},
  \bibinfo{person}{Shreeyash Gowaikar}, \bibinfo{person}{Chandan Gupta},
  \bibinfo{person}{Aman Chadha}, \bibinfo{person}{Aishwarya~Naresh Reganti},
  \bibinfo{person}{Amit Sheth}, {and} \bibinfo{person}{Amitava Das}.}
  \bibinfo{year}{2023}\natexlab{}.
\newblock \showarticletitle{{ANALOGICAL} - {A} {Novel} {Benchmark} for {Long}
  {Text} {Analogy} {Evaluation} in {Large} {Language} {Models}}. In
  \bibinfo{booktitle}{\emph{Findings of the {Association} for {Computational}
  {Linguistics}: {ACL} 2023}}. \bibinfo{publisher}{Association for
  Computational Linguistics}, \bibinfo{address}{Toronto, Canada},
  \bibinfo{pages}{3534--3549}.
\newblock
\urldef\tempurl%
\url{https://doi.org/10.18653/v1/2023.findings-acl.218}
\showDOI{\tempurl}


\bibitem[Yan et~al\mbox{.}(2024)]%
        {yan2024practical}
\bibfield{author}{\bibinfo{person}{Lixiang Yan}, \bibinfo{person}{Lele Sha},
  \bibinfo{person}{Linxuan Zhao}, \bibinfo{person}{Yuheng Li},
  \bibinfo{person}{Roberto Martinez-Maldonado}, \bibinfo{person}{Guanliang
  Chen}, \bibinfo{person}{Xinyu Li}, \bibinfo{person}{Yueqiao Jin}, {and}
  \bibinfo{person}{Dragan Ga{\v{s}}evi{\'c}}.} \bibinfo{year}{2024}\natexlab{}.
\newblock \showarticletitle{Practical and ethical challenges of large language
  models in education: A systematic scoping review}.
\newblock \bibinfo{journal}{\emph{British Journal of Educational Technology}}
  \bibinfo{volume}{55}, \bibinfo{number}{1} (\bibinfo{year}{2024}),
  \bibinfo{pages}{90--112}.
\newblock


\bibitem[Yeo et~al\mbox{.}(2024)]%
        {help2024yeo}
\bibfield{author}{\bibinfo{person}{ShunYi Yeo}, \bibinfo{person}{Gionnieve
  Lim}, \bibinfo{person}{Jie Gao}, \bibinfo{person}{Weiyu Zhang}, {and}
  \bibinfo{person}{Simon~Tangi Perrault}.} \bibinfo{year}{2024}\natexlab{}.
\newblock \showarticletitle{Help Me Reflect: Leveraging Self-Reflection
  Interface Nudges to Enhance Deliberativeness on Online Deliberation
  Platforms}. In \bibinfo{booktitle}{\emph{Proceedings of the CHI Conference on
  Human Factors in Computing Systems}} (Honolulu, HI, USA)
  \emph{(\bibinfo{series}{CHI '24})}. \bibinfo{publisher}{Association for
  Computing Machinery}, \bibinfo{address}{New York, NY, USA}, Article
  \bibinfo{articleno}{806}, \bibinfo{numpages}{32}~pages.
\newblock
\showISBNx{9798400703300}
\urldef\tempurl%
\url{https://doi.org/10.1145/3613904.3642530}
\showDOI{\tempurl}


\bibitem[Yu et~al\mbox{.}(2014a)]%
        {distributed2014yu}
\bibfield{author}{\bibinfo{person}{Lixiu Yu}, \bibinfo{person}{Aniket Kittur},
  {and} \bibinfo{person}{Robert~E. Kraut}.} \bibinfo{year}{2014}\natexlab{a}.
\newblock \showarticletitle{Distributed analogical idea generation: inventing
  with crowds}. In \bibinfo{booktitle}{\emph{Proceedings of the SIGCHI
  Conference on Human Factors in Computing Systems}} (Toronto, Ontario, Canada)
  \emph{(\bibinfo{series}{CHI '14})}. \bibinfo{publisher}{Association for
  Computing Machinery}, \bibinfo{address}{New York, NY, USA},
  \bibinfo{pages}{1245–1254}.
\newblock
\showISBNx{9781450324731}
\urldef\tempurl%
\url{https://doi.org/10.1145/2556288.2557371}
\showDOI{\tempurl}


\bibitem[Yu et~al\mbox{.}(2014b)]%
        {searching2014yu}
\bibfield{author}{\bibinfo{person}{Lixiu Yu}, \bibinfo{person}{Aniket Kittur},
  {and} \bibinfo{person}{Robert~E. Kraut}.} \bibinfo{year}{2014}\natexlab{b}.
\newblock \showarticletitle{Searching for analogical ideas with crowds}. In
  \bibinfo{booktitle}{\emph{Proceedings of the SIGCHI Conference on Human
  Factors in Computing Systems}} (Toronto, Ontario, Canada)
  \emph{(\bibinfo{series}{CHI '14})}. \bibinfo{publisher}{Association for
  Computing Machinery}, \bibinfo{address}{New York, NY, USA},
  \bibinfo{pages}{1225–1234}.
\newblock
\showISBNx{9781450324731}
\urldef\tempurl%
\url{https://doi.org/10.1145/2556288.2557378}
\showDOI{\tempurl}


\bibitem[Yuan et~al\mbox{.}(2023a)]%
        {yuan-etal-2023-distilling}
\bibfield{author}{\bibinfo{person}{Siyu Yuan}, \bibinfo{person}{Jiangjie Chen},
  \bibinfo{person}{Ziquan Fu}, \bibinfo{person}{Xuyang Ge},
  \bibinfo{person}{Soham Shah}, \bibinfo{person}{Charles Jankowski},
  \bibinfo{person}{Yanghua Xiao}, {and} \bibinfo{person}{Deqing Yang}.}
  \bibinfo{year}{2023}\natexlab{a}.
\newblock \showarticletitle{Distilling Script Knowledge from Large Language
  Models for Constrained Language Planning}. In
  \bibinfo{booktitle}{\emph{Proceedings of the 61st Annual Meeting of the
  Association for Computational Linguistics (Volume 1: Long Papers)}},
  \bibfield{editor}{\bibinfo{person}{Anna Rogers}, \bibinfo{person}{Jordan
  Boyd-Graber}, {and} \bibinfo{person}{Naoaki Okazaki}} (Eds.).
  \bibinfo{publisher}{Association for Computational Linguistics},
  \bibinfo{address}{Toronto, Canada}, \bibinfo{pages}{4303--4325}.
\newblock
\urldef\tempurl%
\url{https://doi.org/10.18653/v1/2023.acl-long.236}
\showDOI{\tempurl}


\bibitem[Yuan et~al\mbox{.}(2023b)]%
        {yuan_beneath_2023}
\bibfield{author}{\bibinfo{person}{Siyu Yuan}, \bibinfo{person}{Jiangjie Chen},
  \bibinfo{person}{Xuyang Ge}, \bibinfo{person}{Yanghua Xiao}, {and}
  \bibinfo{person}{Deqing Yang}.} \bibinfo{year}{2023}\natexlab{b}.
\newblock \showarticletitle{Beneath {Surface} {Similarity}: {Large} {Language}
  {Models} {Make} {Reasonable} {Scientific} {Analogies} after {Structure}
  {Abduction}}. In \bibinfo{booktitle}{\emph{Findings of the {Association} for
  {Computational} {Linguistics}: {EMNLP} 2023}},
  \bibfield{editor}{\bibinfo{person}{Houda Bouamor}, \bibinfo{person}{Juan
  Pino}, {and} \bibinfo{person}{Kalika Bali}} (Eds.).
  \bibinfo{publisher}{Association for Computational Linguistics},
  \bibinfo{address}{Singapore}, \bibinfo{pages}{2446--2460}.
\newblock
\urldef\tempurl%
\url{https://doi.org/10.18653/v1/2023.findings-emnlp.160}
\showDOI{\tempurl}


\bibitem[Yuan et~al\mbox{.}(2023c)]%
        {yuan_analogykb_2023}
\bibfield{author}{\bibinfo{person}{Siyu Yuan}, \bibinfo{person}{Jiangjie Chen},
  \bibinfo{person}{Changzhi Sun}, \bibinfo{person}{Jiaqing Liang},
  \bibinfo{person}{Yanghua Xiao}, {and} \bibinfo{person}{Deqing Yang}.}
  \bibinfo{year}{2023}\natexlab{c}.
\newblock \showarticletitle{{ANALOGYKB}: {Unlocking} {Analogical} {Reasoning}
  of {Language} {Models} with {A} {Million}-scale {Knowledge} {Base}}.
\newblock \bibinfo{journal}{\emph{arXiv preprint arXiv:2305.05994}}
  (\bibinfo{year}{2023}).
\newblock


\bibitem[Zhang et~al\mbox{.}(2024)]%
        {cflow}
\bibfield{author}{\bibinfo{person}{Ashley~Ge Zhang}, \bibinfo{person}{Xiaohang
  Tang}, \bibinfo{person}{Steve Oney}, {and} \bibinfo{person}{Yan Chen}.}
  \bibinfo{year}{2024}\natexlab{}.
\newblock \showarticletitle{CFlow: Supporting Semantic Flow Analysis of
  Students' Code in Programming Problems at Scale}. In
  \bibinfo{booktitle}{\emph{Proceedings of the Eleventh ACM Conference on
  Learning @ Scale}} (Atlanta, GA, USA) \emph{(\bibinfo{series}{L@S '24})}.
  \bibinfo{publisher}{Association for Computing Machinery},
  \bibinfo{address}{New York, NY, USA}, \bibinfo{pages}{188–199}.
\newblock
\showISBNx{9798400706332}
\urldef\tempurl%
\url{https://doi.org/10.1145/3657604.3662025}
\showDOI{\tempurl}


\bibitem[Zhu et~al\mbox{.}(2023)]%
        {bilogically2023zhu}
\bibfield{author}{\bibinfo{person}{Qihao Zhu}, \bibinfo{person}{Xinyu Zhang},
  {and} \bibinfo{person}{Jianxi Luo}.} \bibinfo{year}{2023}\natexlab{}.
\newblock \showarticletitle{{Biologically Inspired Design Concept Generation
  Using Generative Pre-Trained Transformers}}.
\newblock \bibinfo{journal}{\emph{Journal of Mechanical Design}}
  \bibinfo{volume}{145}, \bibinfo{number}{4} (\bibinfo{date}{01}
  \bibinfo{year}{2023}), \bibinfo{pages}{041409}.
\newblock
\showISSN{1050-0472}
\urldef\tempurl%
\url{https://doi.org/10.1115/1.4056598}
\showDOI{\tempurl}
\showeprint{https://asmedigitalcollection.asme.org/mechanicaldesign/article-pdf/145/4/041409/6974748/md\_145\_4\_041409.pdf}


\end{thebibliography}

\end{document}